\documentclass[pdflatex,sn-mathphys-num]{sn-jnl}% Math and Physical Sciences Numbered Reference Style
\usepackage{graphicx}%
\usepackage{multirow}%
\usepackage{amsmath,amssymb,amsfonts}%
\usepackage{amsthm}%
\usepackage{rsfso}
\DeclareMathAlphabet{\mathscr}{OMS}{rsfso}{m}{n}
\usepackage[title]{appendix}%
\usepackage{xcolor}%
\usepackage{textcomp}%
\usepackage{manyfoot}%
\usepackage{booktabs}%
\usepackage{algorithm}%
\usepackage{algorithmicx}%
\usepackage{algpseudocode}%
\usepackage{listings}%
\usepackage{url}%
\usepackage{hyperref}%
\hypersetup{bookmarksdepth=3} % include up to \subsubsection in PDF bookmarks (no \paragraph)
\usepackage{tikz}
\usetikzlibrary{calc}
\usepackage{makecell}
\usepackage{booktabs}
\usepackage{tabularx}
\usepackage{threeparttable}
\theoremstyle{thmstyleone}%
\DeclareMathOperator{\softmax}{softmax}
\DeclareMathOperator{\CE}{CE}
\DeclareMathOperator{\KL}{KL}
\DeclareMathOperator{\Tr}{Tr}
\newcommand{\E}{\mathbb{E}}
\newcommand{\R}{\mathbb{R}}

\theoremstyle{thmstyletwo}%

\theoremstyle{thmstylethree}%
\renewcommand{\thefigure}{\arabic{figure}}
\renewcommand{\thetable}{\arabic{table}}
\renewcommand{\thealgorithm}{\arabic{algorithm}}
\begin{document}

\title[Photonic Quantum-Enhanced Knowledge Distillation]{Photonic Quantum-Enhanced Knowledge Distillation}


\author*[1,2,8]{\fnm{Kuan-Cheng} \sur{Chen}}\email{kuan-cheng.chen17@imperial.ac.uk}
\equalcont{These authors contributed equally to this work.}
\author[3,2]{\fnm{Shang} \sur{Yu}}\email{shang.yu@imperial.ac.uk}
\equalcont{These authors contributed equally to this work.}
\author[4]{\fnm{Chen-Yu} \sur{Liu}}\email{: d10245003@g.ntu.edu.tw}
\author[5]{\fnm{Samuel Yen-Chi} \sur{Chen}}\email{ycchen1989@ieee.org}
\author[5]{\fnm{Huan-Hsin} \sur{Tseng}}\email{htseng@bnl.gov}
\author[6,7]{\fnm{Yen Jui} \sur{Chang}}\email{aceest@cycu.edu.tw}
\author[8]{\fnm{Wei-Hao} \sur{Huang}}\email{w.huang@j-ij.com}
\author[1,2]{\fnm{Felix} \sur{Burt}}\email{f.burt23@imperial.ac.uk}
\author[9]{\fnm{Esperanza} \sur{Cuenca Gomez}}\email{ecuencagomez@nvidia.com}
\author[9]{\fnm{Zohim} \sur{Chandani}}\email{zchandani@nvidia.com}
\author[10]{\fnm{William} \sur{Clements}}\email{wclements@orcacomputing.com}
\author[3,2,11]{\fnm{Ian} \sur{Walmsley}}\email{ian.walmsley@physics.ox.ac.uk}
\author[1]{\fnm{Kin K.} \sur{Leung}}\email{kin.leung@imperial.ac.uk}

\affil*[1]{\orgdiv{Department of Electrical and Electronic Engineering}, \orgname{Imperial College London}, \orgaddress{\street{South Kensington}, \city{London}, \postcode{SW7 2AZ}, \state{England}, \country{United Kingdom}}}
\affil[2]{\orgdiv{Imperial Centre for Quantum Engineering, Science and Technology}, \orgname{Imperial College London}, \orgaddress{\street{South Kensington}, \city{London}, \postcode{SW7 2AZ}, \state{England}, \country{United Kingdom}}}
\affil[3]{\orgdiv{Blackett Laboratory, Department of Physics}, \orgname{Imperial College London}, \orgaddress{\street{South Kensington}, \city{London}, \postcode{SW7 2AZ}, \state{England}, \country{United Kingdom}}}
\affil[4]{\orgdiv{Graduate Institute of Applied Physics}, \orgname{National Taiwan University}, \orgaddress{\street{No. 1, Sec. 4, Roosevelt Rd.}, \city{Taipei}, \postcode{106319}, \country{Taiwan}}}
\affil[5]{\orgdiv{Computational Science Initiative}, \orgname{Brookhaven National Laboratory}, \orgaddress{\street{Bldg.\ 725, Room 2-189, P.O.\ Box 5000}, \city{Upton}, \state{NY}, \postcode{11973-5000}, \country{USA}}}
\affil[6]{\orgdiv{Quantum Information Center}, \orgname{Chung Yuan Christian University}, \orgaddress{\street{No.\ 200, Zhongbei Rd., Zhongli Dist.}, \city{Taoyuan City}, \postcode{320314}, \country{Taiwan}}}
\affil[7]{\orgdiv{Master Program in Intelligent Computing and Big Data}, \orgname{Chung Yuan Christian University}, \orgaddress{\street{Zhongbei Rd}, \city{aoyuan City}, \postcode{320314}, \country{Taiwan}}}

\affil[8]{\orgname{JIJ}, \orgaddress{\street{Rutherford Appleton Laboratory, Harwell Campus}, \city{Didcot}, \postcode{OX11 0QX}, \country{United Kingdom}}}

\affil[8]{\orgname{JIJ}, \orgaddress{\street{3-3-6 Shibaura, Minato-ku}, \city{Tokyo}, \postcode{108-0023}, \country{Japan}}}

\affil[9]{\orgname{NVIDIA Corporation}, \orgaddress{\city{Santa Clara}, \state{CA}, \country{USA}}}
\affil[10]{\orgname{ORCA Computing}, \orgaddress{\city{London}, \country{United Kingdom}}}
\affil[11]{\orgdiv{Department of Physics}, \orgname{University of Oxford}, \orgaddress{\street{Parks Road}, \city{Oxford}, \postcode{OX1 3PU}, \state{England}, \country{United Kingdom}}}


\abstract{Photonic quantum processors naturally produce intrinsically stochastic measurement outcomes, offering a hardware-native source of structured randomness that can be exploited during machine-learning training. Here we introduce Photonic Quantum-Enhanced Knowledge Distillation (PQKD), a hybrid quantum photonic--classical framework in which a programmable photonic circuit generates a compact conditioning signal that constrains and guides a parameter-efficient student network during distillation from a high-capacity teacher. PQKD replaces fully trainable convolutional kernels with dictionary convolutions: each layer learns only a small set of shared spatial basis filters, while sample-dependent channel-mixing weights are derived from shot-limited photonic features and mapped through a fixed linear transform. Training alternates between standard gradient-based optimisation of the student and sampling-robust, gradient-free updates of photonic parameters, avoiding differentiation through photonic hardware. Across MNIST, Fashion-MNIST and CIFAR-10, PQKD traces a controllable compression--accuracy frontier, remaining close to teacher performance on simpler benchmarks under aggressive convolutional compression. Performance degrades predictably with finite sampling, consistent with shot-noise scaling, and exponential moving-average feature smoothing suppresses high-frequency shot-noise fluctuations, extending the practical operating regime at moderate shot budgets.}




\maketitle

\section{Introduction}\label{sec1}
As the realm of computing continues to evolve, the integration of optical technologies has emerged as a decisive frontier for processing and information transfer\cite{mcmahon2023physics,solli2015analog}. Traditional electronic computing faces increasing constraints in speed, bandwidth, and energy efficiency as data demands accelerate in an increasingly connected world\cite{waldrop2016chips}. Optical computing, which exploits the physics of light for high-bandwidth transport and parallelizable transformations, offers an alternative path toward ultra-fast data movement and potentially more energy-efficient computation\cite{xue2022nanosecond}. Recent advances in photonic devices—including low-loss waveguides\cite{bose2024anneal}, stable on-chip lasers\cite{xiang2021high}, high-speed modulators\cite{wang2018integrated}, and scalable integrated photonic circuits\cite{shekhar2024roadmapping}—make this an especially timely moment to explore unconventional optical computing architectures that go beyond incremental improvements to electronic logic\cite{hua2025integrated}.

Within this broader landscape, photonic quantum computing provides a distinctive set of computational primitives rooted in quantum interference and measurement\cite{yu2025QTDA,psiquantum2025manufacturable,yu2023GBS,wang2020integrated,bromley2020applications,zhong2020quantum,o2009photonic,knill2001scheme}. 
Of particular relevance to near-term implementations is the continuous-variable (CV) paradigm, where information is encoded in field quadratures and manipulated using Gaussian operations (such as squeezing, phase shifts, and interferometry), optionally augmented by non-Gaussian resources and measurements\cite{weedbrook2012gaussian,braunstein2005quantum,lloyd1999quantum}. 
CV photonic platforms are naturally aligned with integrated optics\cite{stokowski2023integrated} and can operate at high repetition rates\cite{asavanant2021time,asavanant2019generation}, making them compelling candidates for hardware-in-the-loop learning and optimization\cite{cimini2024variational}. 
Recent community efforts have also begun to establish reproducible baselines for photonic quantum machine learning under hardware-feasible constraints\cite{notton2025photonic_baselines}, complementing broader perspectives on optical routes to quantum neural networks\cite{yu2024shedding_light}. 
At the same time, the shot-limited nature of photonic measurements and the practical constraints of near-term devices motivate algorithmic designs that can extract utility from stochastic measurement statistics without requiring fault tolerance or full end-to-end differentiability\cite{preskill2018quantum,scriva2024challenges,katabarwa2024early}.

In parallel, machine learning models—particularly convolutional neural networks—have achieved remarkable performance, yet their growing parameter counts hinder deployment in resource-constrained settings\cite{lecun2015deep,cheng2018model,howard2017mobilenets}. 
Model compression has therefore become central to practical AI, and knowledge distillation is among the most effective strategies: a compact student model learns not only from labelled data but also from a high-capacity teacher’s predictive behaviour\cite{cheng2018model,hinton2015distilling,romero2015experimental}. 
Beyond classical KD, quantum-enhanced conditioning signals and kernelized relations have recently been explored as training-time resources while preserving fully classical deployment\cite{liu2025qrkd,chen2025qsvm_validation,chen2024stellar_qsvm}. 
However, aggressively shrinking a student often degrades performance, especially when compression targets the early convolutional layers that shape representation quality\cite{zeiler2014visualizing,vadera2022methods}. 
This tension suggests an opportunity for new compression mechanisms that retain expressivity while drastically reducing the number of trainable degrees of freedom, particularly when the additional structure is used only during training.

Here we propose Photonic Quantum-Enhanced Knowledge Distillation (PQKD), a hybrid quantum-photonic–classical framework that links photonic CV hardware with parameter-efficient distillation. 
The design is inspired by a line of training-time-only quantum-assisted parameter-efficient learning approaches (e.g., Quantum-Train and its variants), which leverage quantum models to generate or modulate classical parameters during training while keeping inference purely classical\cite{lin2024qtlstm_flood,liu2024qtrl,lin2024qtcnn_deepfake,liu2024qt_tn_mapping,chen2024distqtrl,liu2025qpa_typhoon}. 
PQKD uses a CV photonic quantum circuit as a compact, trainable generator of structured stochastic signals that condition a parameter-efficient student network during distillation. 
Rather than learning full convolutional kernels, the student represents each convolutional layer using a low-dimensional basis, while a photonic feature vector—computed from shot-limited measurement statistics of a 16-mode circuit—controls how those bases mix across channels. 
Training alternates between (i) updating the student with standard distillation objectives to match the teacher and (ii) updating the photonic circuit parameters with sampling-robust, gradient-free optimization compatible with finite-shot measurements and black-box hardware access.

We validate PQKD on MNIST as a controlled benchmark where compression–accuracy trade-offs can be quantified precisely and where the effects of sampling noise can be studied without confounding factors. In this setting, PQKD maintains strong classification performance while enabling substantially stronger compression than conventional students, particularly when compression is applied across the full convolutional stack. The approach exposes clear, experimentally meaningful knobs—such as the basis rank used to parameterize convolutional filters and the photonic parameter budget—that directly determine the achievable compression–accuracy frontier.

More broadly, PQKD illustrates how integrated photonic platforms—especially CV architectures—can contribute to practical learning workflows not only as accelerators for linear optical transforms, but as hardware-aligned generators of rich measurement statistics that can be harnessed to reduce model complexity in AI systems. By bridging photonic quantum hardware constraints with mainstream distillation pipelines, PQKD provides a blueprint for future studies on larger datasets, more expressive optical circuits, and rigorous comparisons against classical conditioning and hypernetwork baselines under matched parameter budgets.

\section{Results}\label{sec2}

\begin{figure}[b]
    \centering
    \includegraphics[width=\linewidth]{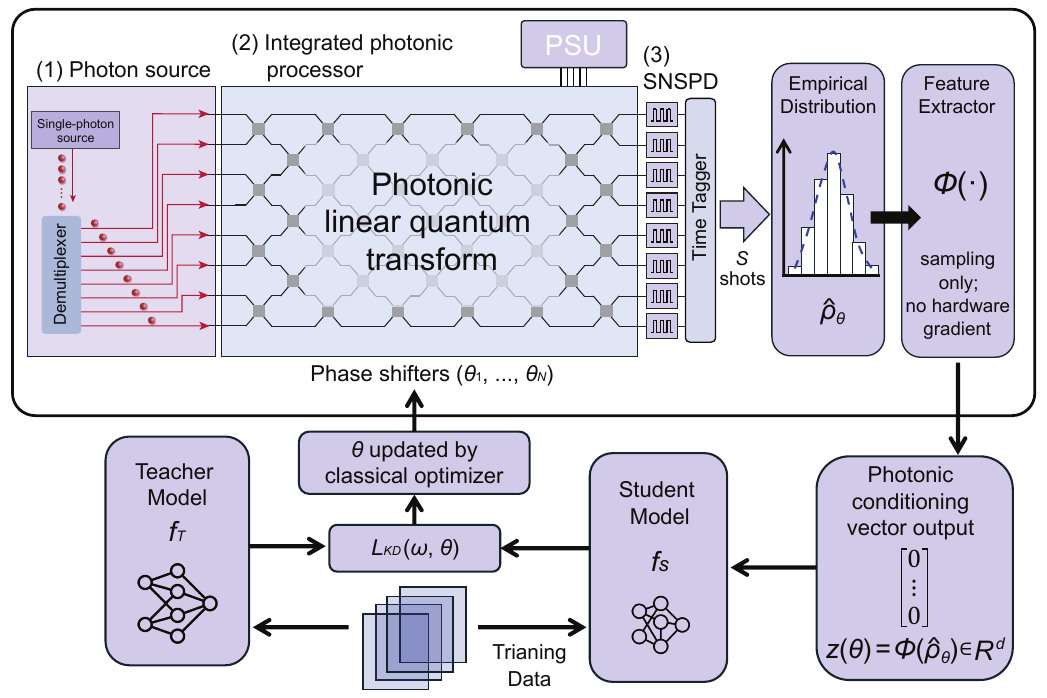}
    \caption{\textbf{Hardware--algorithm co-design for PQKD enabling neural network compression.}
    A CV photonic module prepares a fixed input state \(\rho_{\mathrm{in}}\) and applies a programmable unitary \(U(\theta)\) implemented by an integrated interferometric mesh. The output state \(\rho_\theta = U(\theta)\rho_{\mathrm{in}}U^\dagger(\theta)\) is measured using an SNSPD array with time-tagging, producing i.i.d. samples \(\{\omega_s\}_{s=1}^S\) and an empirical distribution \(\hat{p}_\theta\). A robust classical feature extractor \(\Phi(\cdot)\) maps \(\hat{p}_\theta\) to a fixed-length conditioning vector \(z(\theta)=\Phi(\hat{p}_\theta)\in\mathbb{R}^{d}\), which modulates the student model \(f_S\) during knowledge distillation from a pretrained teacher \(f_T\). Crucially, \(\theta\) is updated by a classical optimiser using a distillation objective \(L_{\mathrm{KD}}(\omega,\theta)\) and requires sampling only, avoiding differentiation through photonic hardware.}
    \label{fig:pqkd_hardware}
\end{figure}

\subsection{Photonic quantum-enhanced knowledge distillation as a structured compression mechanism}
\label{sec:results_theory}

\paragraph{Teacher--student distillation objective.}

Let $\mathcal{D}=\{(x_n,y_n)\}_{n=1}^N$ denote the training set with $y_n\in\{1,\dots,C\}$. A trained teacher network produces logits $t(x)\in\mathbb{R}^C$ and the student produces logits $s(x;w,\theta)\in\mathbb{R}^C$, where $w$ denotes trainable \emph{classical} parameters in the student and $\theta$ denotes \emph{photonic circuit} parameters. For temperature $\tau>0$, define softened distributions
\begin{equation}
p_T^{(\tau)}(x)=\mathrm{softmax}\!\left(\frac{t(x)}{\tau}\right),\qquad
p_S^{(\tau)}(x;w,\theta)=\mathrm{softmax}\!\left(\frac{s(x;w,\theta)}{\tau}\right).
\label{eq:soft_probs}
\end{equation}

We train the student using a standard distillation objective combining hard-label supervision and teacher matching [See Supplementary Information]:
\begin{equation}
\mathcal{L}_{\mathrm{KD}}(w,\theta)=
\mathbb{E}_{(x,y)\sim \mathcal{D}}
\Big[
\lambda\,\mathrm{CE}\!\left(y,\,p_S^{(1)}(x;w,\theta)\right)
+
(1-\lambda)\,\tau^2\,\mathrm{KL}\!\left(p_T^{(\tau)}(x)\,\Vert\,p_S^{(\tau)}(x;w,\theta)\right)
\Big],
\label{eq:kd_loss}
\end{equation}
with $\lambda\in[0,1]$. The KL term is equivalent (up to an additive constant independent of student parameters) to cross-entropy against teacher soft targets:
\begin{equation}
\mathrm{KL}\!\left(p_T^{(\tau)}\Vert p_S^{(\tau)}\right)
=
\mathrm{CE}\!\left(p_T^{(\tau)},p_S^{(\tau)}\right)-\mathrm{H}\!\left(p_T^{(\tau)}\right),
\label{eq:kl_ce_relation}
\end{equation}
thereby transferring the teacher's \emph{relative class structure} to the student. This is particularly important under aggressive compression, where hard-label training alone can be insufficient to identify a well-generalising solution within a restricted hypothesis class.

\paragraph{Continuous-variable photonic quantum feature map.}

PQKD couples distillation to a photonic module that produces a compact, distributional conditioning signal. Let $\rho_{\mathrm{in}}$ be a fixed input state over $N$ photonic modes and let $U(\theta)$ be a parameterised CV unitary realised by an integrated photonic circuit (e.g., interferometric networks with phase shifts). The circuit output state is
\begin{equation}
\rho_{\theta}=U(\theta)\,\rho_{\mathrm{in}}\,U(\theta)^{\dagger}.
\label{eq:rho_theta}
\end{equation}

Let $\Omega$ denote the (finite or countable) set of classical measurement outcomes and let
$\{M_{\omega}\}_{\omega\in\Omega}$ be a POVM on the $N$-mode Hilbert space (covering projective measurements as a special case,
as well as realistic detectors such as photon-number-resolving or threshold detection). This induces the output distribution
\begin{equation}
p_{\theta}(\omega)=\mathrm{Tr}\!\left(M_{\omega}\,\rho_{\theta}\right),\qquad \omega\in\Omega,\qquad \sum_{\omega\in\Omega}p_{\theta}(\omega)=1.
\label{eq:photonic_distribution}
\end{equation}
With $S$ shots, we obtain i.i.d.\ outcomes $\omega_1,\ldots,\omega_S\sim p_{\theta}$ and an empirical distribution $\widehat{p}_{\theta}$.
The photonic feature is a deterministic, fixed-length summary
\begin{equation}
z(\theta)=\Phi(\widehat{p}_{\theta})\in\mathbb{R}^{d},
\label{eq:z_theta}
\end{equation}
where in our implementation $N=16$ and $d=512$ (constructed from stable marginals of the measurement statistics; see Methods).
Notably, PQKD does not require differentiating through the photonic hardware; it requires only sampling and a robust feature extractor $\Phi$.

\paragraph{Compressed convolution as a photonic-conditioned dictionary model.}
We consider a \emph{convolutional neural network (CNN)} layer whose learnable weights are a 4th-order kernel tensor
$W\in\mathbb{R}^{C_{\mathrm{out}}\times C_{\mathrm{in}}\times k\times k}$ and bias
$b\in\mathbb{R}^{C_{\mathrm{out}}}$.
Here $C_{\mathrm{in}}$ and $C_{\mathrm{out}}$ denote the numbers of input and output channels, and
$k\times k$ is the spatial kernel size.

PQKD replaces the fully trainable tensor $W$ with a structured parameterisation that separates
\emph{spatial structure} from \emph{channel mixing}.
Specifically, we introduce $R$ trainable spatial basis filters
$B\in\mathbb{R}^{R\times k\times k}$ and a channel-mixing tensor
$M\in\mathbb{R}^{C_{\mathrm{out}}\times C_{\mathrm{in}}\times R}$ such that, for each output channel $o$,
input channel $i$, and spatial indices $\alpha,\beta\in\{1,\ldots,k\}$,
\begin{equation}
W_{o, i, \alpha, \beta}
=\sum_{r=1}^{R} M_{o,i,r}\,B_{r, \alpha, \beta}.
\label{eq:kernel_recon}
\end{equation}
Equivalently, in tensor form one may write $W_{o,i,:,:}=\sum_{r=1}^{R} M_{o,i,r}\,B_{r,:,:}$, but
Eq.~\eqref{eq:kernel_recon} makes the index structure explicit.

To avoid learning the large tensor $M$ directly, PQKD constrains it to be generated from the photonic feature $z(\theta)$ via a fixed linear map:
\begin{equation}
\mathrm{vec}(M)=A\,z(\theta),
\label{eq:mixing_linear}
\end{equation}
where $A\in\mathbb{R}^{(C_{\mathrm{out}}C_{\mathrm{in}}R)\times d}$ is fixed at initialisation (Methods). Equations~\eqref{eq:kernel_recon}--\eqref{eq:mixing_linear} imply that the student convolutional kernels lie on a low-dimensional manifold controlled by $(B,\theta)$.

\paragraph{Parameter-count reduction.}
For a compressed convolutional layer, the number of trainable parameters is $Rk^2+C_{\mathrm{out}}$ for $(B,b)$, plus the \emph{global} photonic parameter budget $\dim(\theta)$ (shared across layers in our current design). In contrast, a dense kernel contains $C_{\mathrm{out}}C_{\mathrm{in}}k^2+C_{\mathrm{out}}$ trainable parameters. Ignoring the shared $\theta$ term, the approximate per-layer compression factor is therefore
\begin{equation}
\mathrm{CR}_{\mathrm{conv}}
\approx
\frac{C_{\mathrm{out}}C_{\mathrm{in}}k^2}{Rk^2+C_{\mathrm{out}}}.
\label{eq:compression_ratio}
\end{equation}
When compressing multiple convolutional layers, the dominant trainable budget scales with $\sum_{\ell} R_{\ell}k_{\ell}^2$ rather than $\sum_{\ell} C_{\mathrm{out}}^{(\ell)}C_{\mathrm{in}}^{(\ell)}k_{\ell}^2$, enabling substantially larger end-to-end compression when the scheme is applied to the full convolutional stack.

\paragraph{Bilevel interpretation of photonic knowledge distillation.}
PQKD optimises the student weights $w$ and photonic parameters $\theta$ under a coupled objective. A natural formulation is bilevel optimisation:
\begin{equation}
w^{\ast}(\theta)\in \arg\min_{w}\ \mathcal{L}^{\mathrm{train}}_{\mathrm{KD}}(w,\theta),
\qquad
\theta^{\ast}\in \arg\min_{\theta}\ \mathcal{L}^{\mathrm{val}}_{\mathrm{KD}}(w^{\ast}(\theta),\theta),
\label{eq:bilevel-main}
\end{equation}
where the outer objective uses a validation proxy that is robust to shot noise. Our training loop approximates Eq.~\eqref{eq:bilevel-main} by alternating (i) gradient-based updates for $w$ and (ii) sampling-robust, gradient-free updates for $\theta$ (Methods). In this view, the teacher supervises the student through softened logits, while the photonic module \emph{adapts the restricted student hypothesis class} by shaping $z(\theta)$, thereby selecting a compressed parameter manifold that best matches the teacher.

\paragraph{Finite-shot stability and noise propagation.}
Shot-limited sampling produces $\widehat{p}_{\theta}$ and hence a noisy feature $\widehat{z}(\theta)$. For each bin $\omega$, Hoeffding's inequality gives
\begin{equation}
\mathbb{P}\!\left(\left|\widehat{p}_{\theta}(\omega)-p_{\theta}(\omega)\right|\ge \epsilon\right)\le 2\exp(-2S\epsilon^2).
\label{eq:hoeffding}
\end{equation}
If $\Phi$ is Lipschitz in the histogram (true for the marginal-based mapping used here away from degenerate normalisation; Methods), then with high probability
\begin{equation}
\|\widehat{z}(\theta)-z(\theta)\|_2 \le L_{\Phi}\,\|\widehat{p}_{\theta}-p_{\theta}\|_1 = \mathcal{O}\!\left(\sqrt{\frac{K}{S}}\right),
\label{eq:feature_concentration}
\end{equation}
where $K=|\Omega|$ is the effective number of histogram bins. Because $M$ depends linearly on $z$ and $W$ is linear in $M$, the reconstructed kernel is Lipschitz in $z$:
\begin{equation}
\|W(B,\widehat{z})-W(B,z)\|_F \le \|A\|_2\,\|B\|_F\,\|\widehat{z}-z\|_2,
\label{eq:kernel_lipschitz}
\end{equation}
and therefore weight and logit perturbations decay with $S$ in a controlled manner. This provides a theoretical basis for the observed smooth degradation under reduced shot budgets and motivates feature smoothing (e.g., exponential moving averages (EMA)) as a variance-reduction mechanism during training.

\subsection{Training and validation dynamics}
\label{sec:exp1}

We evaluate PQKD on MNIST, Fashion-MNIST and CIFAR-10 using a unified experimental protocol (Methods; Supplementary Information). The teacher is a three-layer convolutional network, and the PQKD student preserves the same overall structure but compresses the first two convolutional layers by restricting their kernels to a low-dimensional manifold generated by a shot-limited photonic module. This reduces the effective parameter count from \(\sim 2.25\times10^{5}\) (teacher) to \(\sim 1.50\times10^{5}\) (student), while introducing only 45 trainable photonic parameters \(\theta\), corresponding to \(\sim\!1.5\times\) compression. We focus here on partial convolutional compression to isolate training and generalisation dynamics; full-network compression and scaling results are presented in the next subsection.

\begin{table*}[b]
\centering
\scriptsize
\setlength{\tabcolsep}{3.0pt}
\renewcommand{\arraystretch}{1.12}
\newcommand{\pmcell}[2]{\shortstack{#1\\($\pm$ #2)}}

\caption{\textbf{PQKD performance and optimisation (Opt.) diagnostics.}\\
Values are mean $\pm$ s.d. over five independent runs. Accuracies are in \%.
CE denotes cross-entropy. Photonic $\Delta$ is (epoch 1)$-$(final epoch).}
\label{tab:pqkd_summary}

\begin{tabularx}{\textwidth}{l *{7}{>{\centering\arraybackslash}X}}
\toprule
\textbf{Dataset} &
\multicolumn{4}{c}{\textbf{Accuracy (\%)}} &
\multicolumn{2}{c}{\textbf{Final CE loss}} &
\textbf{Opt.} \\
\cmidrule(lr){2-5}\cmidrule(lr){6-7}\cmidrule(lr){8-8}

% ---- Row 2: train/val first ----
&
\multicolumn{2}{c}{\textbf{train}} &
\multicolumn{2}{c}{\textbf{validation}} &
\multicolumn{2}{c}{\textbf{validation}} &
\textbf{Photonic} \\

% ---- Row 3: Teacher/PQKD second ----
&
\textbf{Teacher} & \textbf{PQKD} &
\textbf{Teacher} & \textbf{PQKD} &
\textbf{Teacher} & \textbf{PQKD} &
\textbf{$\Delta$} \\
\midrule

MNIST &
\textbf{\pmcell{100.00}{0.00}} &
\pmcell{99.71}{0.02} &
\pmcell{99.07}{0.13} &
\textbf{\pmcell{99.09}{0.09}} &
\pmcell{0.031}{0.002} &
\textbf{\pmcell{0.030}{0.001}} &
\pmcell{5.94}{0.02} \\
\addlinespace[2pt]
\cmidrule(lr){1-8}
\addlinespace[2pt]

Fashion-MNIST &
\textbf{\pmcell{100.00}{0.00}} &
\pmcell{97.10}{0.18} &
\pmcell{91.86}{0.26} &
\textbf{\pmcell{92.42}{0.16}} &
\pmcell{0.291}{0.003} &
\textbf{\pmcell{0.239}{0.004}} &
\pmcell{6.73}{0.03} \\
\addlinespace[2pt]
\cmidrule(lr){1-8}
\addlinespace[2pt]

CIFAR-10 &
\textbf{\pmcell{91.81}{0.13}} &
\pmcell{80.35}{0.26} &
\textbf{\pmcell{86.99}{0.29}} &
\pmcell{79.72}{0.37} &
\textbf{\pmcell{0.397}{0.004}} &
\pmcell{0.597}{0.008} &
\pmcell{4.95}{0.02} \\
\bottomrule
\end{tabularx}
\end{table*}

%\begin{figure*}[b]
%\centering

% ==================== Row 1: Train accuracy ====================
%\begin{subfigure}[t]{0.3\textwidth}
%\centering
%\includegraphics[width=\linewidth]{figures/train_acc_a_mnist.pdf}
%\end{subfigure}\hfill
%\begin{subfigure}[t]{0.3\textwidth}
%\centering
%\includegraphics[width=\linewidth]{figures/train_acc_b_fashion.pdf}
%\end{subfigure}\hfill
%\begin{subfigure}[t]{0.3\textwidth}
%\centering
%\includegraphics[width=\linewidth]{figures/train_acc_c_cifar10.pdf}
%\end{subfigure}

%\vspace{0.55em}

% ==================== Row 2: Val accuracy ====================
%\begin{subfigure}[t]{0.3\textwidth}
%\centering
%\includegraphics[width=\linewidth]{figures/fig_a_mnist.pdf}
%\end{subfigure}\hfill
%\begin{subfigure}[t]{0.3\textwidth}
%\centering
%\includegraphics[width=\linewidth]{figures/fig_b_fashion.pdf}
%\end{subfigure}\hfill
%\begin{subfigure}[t]{0.3\textwidth}
%\centering
%\includegraphics[width=\linewidth]{figures/fig_c_cifar10.pdf}
%\end{subfigure}

%\vspace{0.55em}

% ==================== Row 3: CE loss (train+val together) ====================
%\begin{subfigure}[t]{0.3\textwidth}
%\centering
%\includegraphics[width=\linewidth]{figures/ce_loss_train_val_a_mnist.pdf}
%\end{subfigure}\hfill
%\begin{subfigure}[t]{0.3\textwidth}
%\centering
%\includegraphics[width=\linewidth]{figures/ce_loss_train_val_b_fashion.pdf}
%\end{subfigure}\hfill
%\begin{subfigure}[t]{0.3\textwidth}
%\centering
%\includegraphics[width=\linewidth]{figures/ce_loss_train_val_c_cifar10.pdf}
%\end{subfigure}

\vspace{0.35em}

% ==================== Row 4: Dataset labels ====================
%\begin{subfigure}[t]{0.32\textwidth}
%\centering
%\caption{MNIST.}
%\end{subfigure}\hfill
%\begin{subfigure}[t]{0.32\textwidth}
%\centering
%\caption{Fashion-MNIST.}
%\end{subfigure}\hfill
%\begin{subfigure}[t]{0.32\textwidth}
%\centering
%\caption{CIFAR-10.}
%\end{subfigure}

%\caption{\textbf{Training and validation dynamics across datasets.} Mean $\pm$ s.d. over five independent runs comparing the teacher (green) and PQKD student (blue) on MNIST (a), Fashion-MNIST (b), and CIFAR-10 (c). Top row: training accuracy. Middle row: validation accuracy. Bottom row: cross-entropy (CE) loss for training (solid) and validation (dashed)}
%\label{fig:train_curves}
%\end{figure*}

\begin{figure*}[hbt]
\centering
\includegraphics[width=1.00\textwidth]{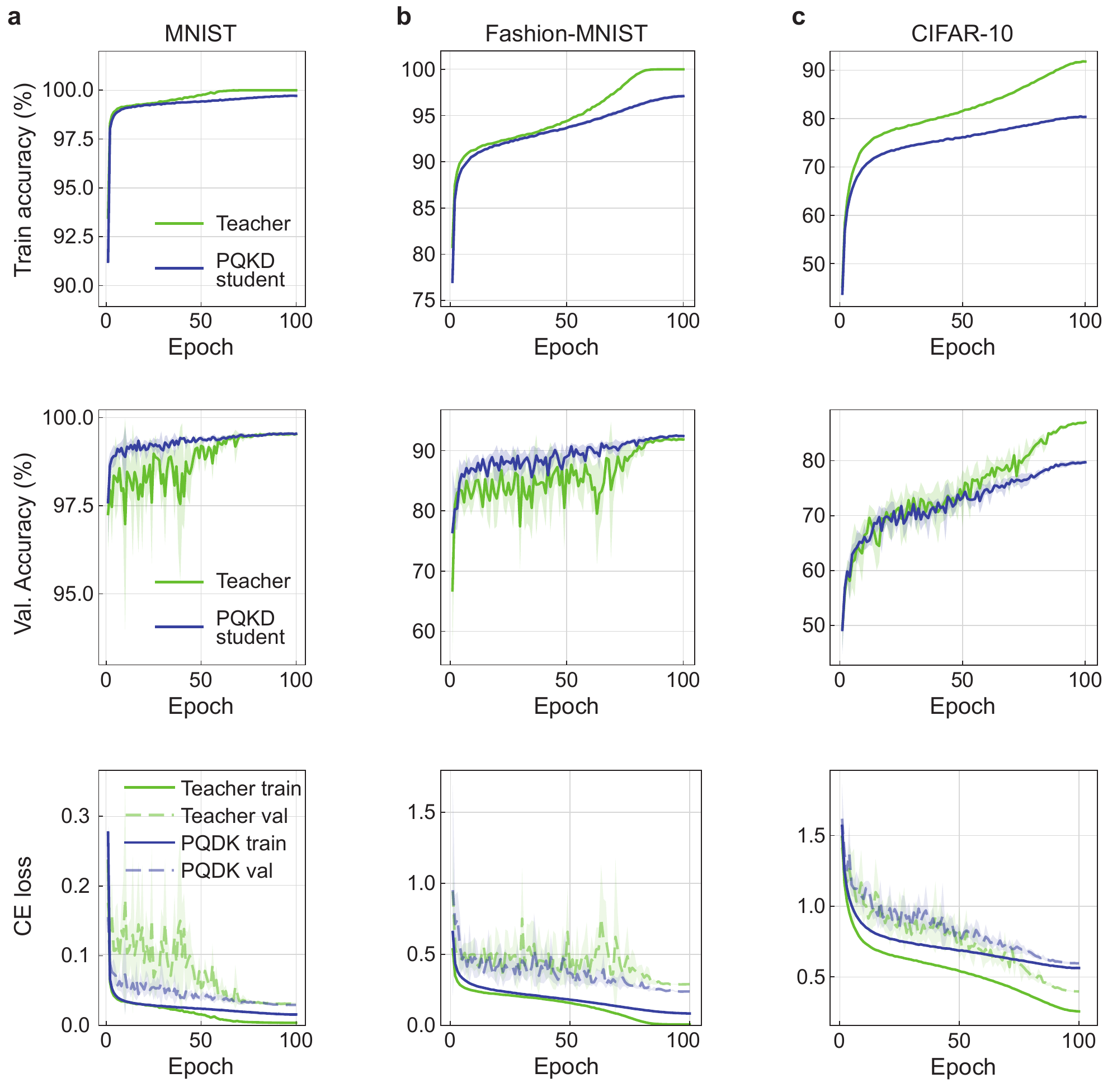}
\caption{\textbf{Training and validation dynamics across datasets.}
Mean $\pm$ s.d. over five independent runs comparing the teacher (green) and PQKD student (blue) on MNIST \textbf{a} , Fashion-MNIST \textbf{b}, and CIFAR-10 \textbf{c}. Top row: training accuracy. Middle row: validation accuracy. Bottom row: cross-entropy (CE) loss for training (solid) and validation (dashed)
}
\label{fig:train_curves}
\end{figure*}

Training and validation dynamics are summarised in Fig.~\ref{fig:train_curves} (mean \(\pm\) s.d. over five independent runs), with aggregate metrics reported in Table~\ref{tab:pqkd_summary}. On MNIST and Fashion-MNIST, the teacher rapidly reaches near-saturated training accuracy, yet exhibits comparatively larger run-to-run fluctuations in validation accuracy and validation CE loss, consistent with over-parameterisation for these tasks. PQKD mitigates this behaviour: under a substantially reduced parameter budget, the student maintains competitive training performance while improving validation stability and achieving favourable validation CE compared with the teacher (Table~\ref{tab:pqkd_summary}). These trends indicate that distillation, combined with the compressed photonic kernel parameterisation, regularises the effective hypothesis space and improves generalisation beyond what is captured by accuracy alone.

For CIFAR-10, the teacher remains stronger in both training and validation accuracy, reflecting increased task complexity under the same compression budget. Nevertheless, PQKD optimisation remains well behaved across datasets: the photonic optimisation diagnostic \(\Delta\) is consistently positive (shown in Table~\ref{tab:pqkd_summary}), indicating systematic progress of the photonic module during training. Overall, Fig.~\ref{fig:train_curves} and Table~\ref{tab:pqkd_summary} show that PQKD can reduce model size while stabilising validation behaviour in regimes where an uncompressed network exhibits signatures of over-fitting.

\subsection{Scaling compression from partial to full convolutional compression}
\label{sec:exp2}

\begin{figure*}[!b]
\centering
\includegraphics[width=\textwidth]{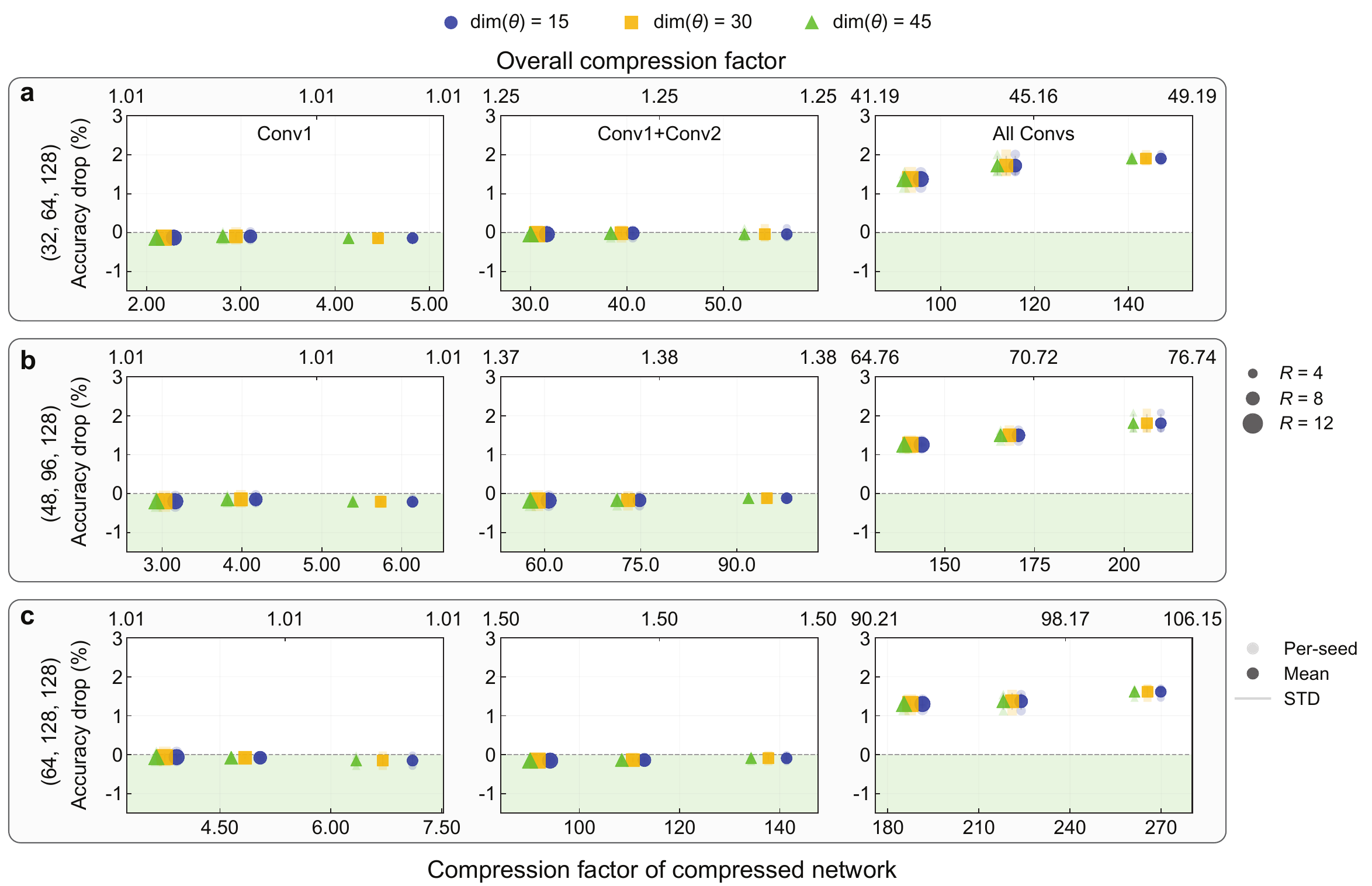}
\caption{\textbf{Scaling photonic-conditioned convolutional compression across network scope and teacher capacity on MNIST.}
Validation accuracy drop relative to the corresponding teacher in percentage points, plotted against the overall compression factor of the compressed network (bottom axis; $\texttt{compression\_x}$). Columns report the compression scope: \textbf{Conv1} (left), \textbf{Conv1+Conv2} (middle), and \textbf{All Convs} (right). Rows correspond to three teacher widths, $(c_1,c_2,c_3)\in\{(32,64,128),(48,96,128),(64,128,128)\}$, with each row aggregating all configurations for that teacher. Marker shape encodes the photonic parameter dimension $\dim(\theta)\in\{15,30,45\}$, and marker size encodes the kernel-manifold rank $R\in\{4,8,12\}$. Faint markers show individual seeds, while opaque markers and vertical bars indicate the mean and one standard deviation across seeds.}
\label{fig:pareto}
\end{figure*}

To assess whether PQKD remains effective under increasingly aggressive structural constraints, we progressively expanded the scope of convolutional compression from early to deep network stages. Specifically, we considered three regimes: compressing only the first convolutional layer (\textbf{Conv1}), compressing the first two convolutional layers (\textbf{Conv1+Conv2}), and compressing the entire convolutional stack (\textbf{Conv1--Conv3}). This controlled progression isolates how far the structured parameterisation defined in Eqs.~\eqref{eq:kernel_recon}--\eqref{eq:mixing_linear} can be pushed before accuracy degrades, and whether knowledge distillation can stabilise learning on increasingly low-dimensional kernel manifolds.

Figure~\ref{fig:pareto} summarises the resulting compression--accuracy landscapes across compression scopes and teacher capacities. Each column corresponds to a compression scope, while rows aggregate results for three teacher widths, $(c_1,c_2,c_3)\in\{(32,64,128),(48,96,128),(64,128,128)\}$. The vertical axis reports the test accuracy drop relative to the corresponding teacher, while the horizontal axis reports the compression factor restricted to the compressed subnetwork. The overall compression factor of the full model is shown on the top axis for reference. Marker shape encodes the photonic parameter budget $\dim(\theta)$ and marker size encodes the basis rank $R$, enabling the joint effects of spatial expressivity and photonic conditioning capacity to be visualised.

\begin{table*}[!t]
\centering
\scriptsize
\setlength{\tabcolsep}{3.0pt}
\renewcommand{\arraystretch}{1.12}
\newcommand{\pmcell}[2]{\shortstack{#1\\($\pm$ #2)}}

\caption{\textbf{Representative operating points for scaling convolutional compression on MNIST (fixed $R{=}4$, $\dim(\theta){=}30$).}\\
$C_x$ denotes the overall compression factor (teacher trainable parameters divided by student trainable parameters, including $\theta$). Accuracies are mean $\pm$ s.d. over 5 seeds, reported at the final epoch.}
\label{tab:mnist_compression_operating_points}

\begin{tabularx}{\textwidth}{l l c *{4}{>{\centering\arraybackslash}X}}
\toprule
\textbf{Scope} &
\shortstack{\textbf{Teacher}\\\textbf{$(c_1,c_2,c_3)$}} &
\textbf{$C_x$} &
\multicolumn{2}{c}{\shortstack{\textbf{Train}\\\textbf{accuracy (\%)}}} &
\multicolumn{2}{c}{\shortstack{\textbf{Validation}\\\textbf{accuracy (\%)}}} \\
\cmidrule(lr){4-5}\cmidrule(lr){6-7}
& & &
\textbf{Teacher} & \textbf{PQKD} &
\textbf{Teacher} & \textbf{PQKD} \\
\midrule

% ---------------- Conv1 ----------------
\multirow{3}{*}{\textbf{Conv1}} &
$(32,64,128)$  & 1.01 &
\textbf{\pmcell{100.00}{0.00}} & \pmcell{99.91}{0.02} &
\pmcell{98.91}{0.09} & \textbf{\pmcell{99.03}{0.08}} \\
& $(48,96,128)$  & 1.01 &
\textbf{\pmcell{100.00}{0.00}} & \pmcell{99.98}{0.00} &
\pmcell{99.11}{0.06} & \textbf{\pmcell{99.20}{0.11}} \\
& $(64,128,128)$ & 1.01 &
\textbf{\pmcell{100.00}{0.00}} & \pmcell{100.00}{0.00} &
\pmcell{99.06}{0.18} & \textbf{\pmcell{99.07}{0.09}} \\

\addlinespace[2pt]
\cmidrule(lr){1-7}
\addlinespace[2pt]

% ---------------- Conv1+2 ----------------
\multirow{3}{*}{\textbf{Conv1+2}} &
$(32,64,128)$  & 1.25 &
\textbf{\pmcell{100.00}{0.00}} & \pmcell{99.45}{0.03} &
\pmcell{98.91}{0.09} & \textbf{\pmcell{98.93}{0.04}} \\
& $(48,96,128)$  & 1.38 &
\textbf{\pmcell{100.00}{0.00}} & \pmcell{99.57}{0.03} &
\textbf{\pmcell{99.11}{0.06}} & \pmcell{99.00}{0.05} \\
& $(64,128,128)$ & 1.50 &
\textbf{\pmcell{100.00}{0.00}} & \pmcell{99.67}{0.04} &
\pmcell{99.06}{0.18} & \textbf{\pmcell{99.10}{0.02}} \\

\addlinespace[2pt]
\cmidrule(lr){1-7}
\addlinespace[2pt]

% ---------------- All Convs ----------------
\multirow{3}{*}{\textbf{All Convs}} &
$(32,64,128)$  & 48.81 &
\textbf{\pmcell{100.00}{0.00}} & \pmcell{96.62}{0.09} &
\textbf{\pmcell{98.91}{0.09}} & \pmcell{96.85}{0.15} \\
& $(48,96,128)$  & 76.18 &
\textbf{\pmcell{100.00}{0.00}} & \pmcell{96.96}{0.17} &
\textbf{\pmcell{99.11}{0.06}} & \pmcell{97.17}{0.03} \\
& $(64,128,128)$ & 105.40 &
\textbf{\pmcell{100.00}{0.00}} & \pmcell{97.19}{0.08} &
\textbf{\pmcell{99.06}{0.18}} & \pmcell{97.33}{0.11} \\

\bottomrule
\end{tabularx}
\end{table*}

When compressing only \textbf{Conv1}, the end-to-end reduction in trainable parameters is modest, as deeper convolutional layers and the classifier continue to dominate the student capacity. Nevertheless, across all teacher widths, the PQKD student closely matches---and in some cases marginally exceeds---the teacher's validation accuracy (Table~\ref{tab:mnist_compression_operating_points}), with negligible accuracy drop. This indicates that early-stage feature extraction can be represented efficiently by a low-rank photonic-conditioned basis without loss of discriminative power, and that the remaining dense layers are sufficient to absorb residual modelling capacity.

Extending compression to \textbf{Conv1+Conv2} substantially increases the fraction of parameters subject to structured constraint, leading to clearer reductions in trainable parameters while preserving strong performance. In this regime, the compression--accuracy trade-off becomes explicit and is jointly controlled by the basis rank $R$ and the photonic parameter budget $\dim(\theta)$, which together determine the dimensionality of the admissible kernel manifold. Across all teacher widths, PQKD maintains validation accuracy within $\sim$0.2--0.3\% of the teacher at overall compression factors of $C_x\approx 1.25$--$1.50$, demonstrating that photonic conditioning and distillation can stabilise learning even when multiple convolutional stages are constrained.

Compressing \textbf{all convolutional layers} yields the largest parameter reductions and exposes a clear compression frontier. In this regime, decreasing the basis rank $R$ increases compression at the cost of higher accuracy drop, while increasing $R$ recovers accuracy with diminishing returns in parameter count. Importantly, even at extreme compression factors exceeding two orders of magnitude in the convolutional subnetwork, PQKD retains stable training and converges to well-defined operating points, rather than exhibiting catastrophic degradation. This behaviour indicates that the combination of photonic-conditioned structured parameterisation and knowledge distillation effectively regularises the optimisation landscape under severe dimensionality reduction.

Representative operating points for each compression scope and teacher width are reported in Table~\ref{tab:mnist_compression_operating_points}. Together, these results demonstrate that PQKD supports a smooth and controllable transition from partial to full convolutional compression, enabling substantial reductions in trainable parameters while preserving accuracy across a wide range of model capacities.

\subsection{Robustness to shot noise and hardware-relevant perturbations}
\label{sec:exp3}

Finally, we probe robustness to stochasticity intrinsic to near-term photonic sampling and to perturbations that capture hardware-relevant variability. In PQKD, the photonic module is queried with a finite measurement budget $S$ to produce an empirical histogram $\widehat{p}_{\theta}$, which is then mapped deterministically to a conditioning feature $\widehat{z}(\theta)=\Phi(\widehat{p}_{\theta})$ (Methods; Supplementary Note~S5). Finite-shot sampling therefore induces a stochastic feature perturbation $\widehat{z}(\theta)-z(\theta)$, where $z(\theta)=\Phi(p_{\theta})$ is the infinite-shot feature. Under a Lipschitz assumption for the histogram-to-feature map (Eq.~\eqref{eq:kernel_lipschitz_supp}), standard multinomial concentration implies
\begin{equation}
\mathbb{E}\!\left[\left\lVert \widehat{z}(\theta)-z(\theta)\right\rVert_2\right]
= \mathcal{O}\!\left(L_{\Phi}\sqrt{\frac{K}{S}}\right),
\label{eq:exp3_feature_shot_scaling}
\end{equation}
where $K$ is the effective number of histogram bins used by $\Phi$ (here $K{=}512$ for two concatenated 256-bin marginals). Equation~\eqref{eq:exp3_feature_shot_scaling} predicts a characteristic $1/\sqrt{S}$ scaling of the shot-induced feature noise, and hence a saturation of performance as $S$ increases.

\begin{figure*}[!t]
\centering
\includegraphics[width=0.8\textwidth]{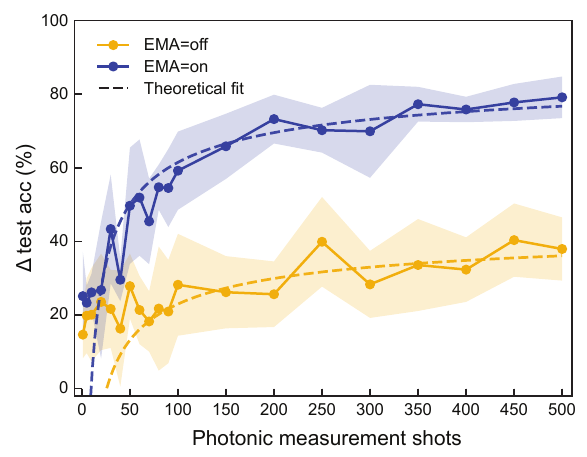}
\caption{\textbf{Shot-limited scaling of the photonic conditioning signal in PQKD.}
Increase in test accuracy due to the photonic feature, $\delta\,\mathrm{test\ acc} \equiv \mathrm{acc}(z)-\mathrm{acc}(z{=}0)$, as a function of the photonic measurement budget $S$ (shots), comparing EMA aggregation of the feature across epochs (blue) with no EMA (orange). Solid curves show the mean over random seeds and shaded bands denote $\pm 1$ s.d. The dashed curves are fits to the shot-noise model $\delta(S)=\delta_\infty-k/\sqrt{S}$, consistent with multinomial histogram fluctuations propagating through a Lipschitz feature map, which predicts a $1/\sqrt{S}$ decay of stochastic feature perturbations and a corresponding saturation of performance at large $S$.}
\label{fig:exp3}
\end{figure*}

To test this prediction, we sweep the shot budget $S$ used to form the photonic conditioning signal during training and evaluation, while holding the student architecture and optimisation schedule fixed. We additionally include two perturbation families that emulate hardware variability: additive feature corruption $\widehat{z}\leftarrow \widehat{z}+\eta$ with $\eta\sim\mathcal{N}(0,\sigma_z^2 I)$, and mild parameter drift $\theta\leftarrow\theta+\xi$ between epochs with $\xi\sim\mathcal{N}(0,\sigma_\theta^2 I)$ (Methods). For each condition we also compare a simple variance-reduction mechanism, exponential moving average (EMA) smoothing of the photonic feature across epochs, which is expected to attenuate high-frequency fluctuations in $\widehat{z}$.

Figure~\ref{fig:exp3} reports an \emph{ablation} that isolates the contribution of the photonic feature by subtracting a matched ``no-photonic'' baseline:
\(
\delta\,\mathrm{test\ acc}\equiv \mathrm{acc}(z)-\mathrm{acc}(z{=}0),
\)
plotted as a function of the shot budget $S$.
The dependence on $S$ is well captured by the theory-motivated form
\begin{equation}
\delta(S)=\delta_{\infty}-\frac{k}{\sqrt{S}},
\label{eq:exp3_delta_model}
\end{equation}
consistent with shot-noise-induced histogram fluctuations propagating through a Lipschitz map $\Phi$.
Weighted nonlinear least-squares fits (using empirical standard deviations as weights; fit restricted to $S\ge 75$) yield
$\delta_{\infty}=46.68\pm10.86$ and $k=236.89\pm160.35$ for EMA-off ($R^2_{w}=0.596$),
and $\delta_{\infty}=89.05\pm5.61$ and $k=275.75\pm85.88$ for EMA-on ($R^2_{w}=0.901$).
The substantially improved goodness-of-fit and higher asymptotic gain under EMA indicate that feature smoothing enlarges the effective operating regime at low-to-moderate shot budgets, mitigating shot-limited fluctuations in the conditioning signal without altering the student architecture. Together, these results support the central prediction of Eq.~\eqref{eq:exp3_feature_shot_scaling}: photonic sampling noise induces a controlled, shot-governed perturbation whose impact can be reduced by simple, hardware-compatible temporal averaging.

\section{Conclusion}
PQKD shows that finite-shot continuous-variable photonic processors can be used as training-time resources for practical model compression. By coupling a programmable photonic sampler to photonic-conditioned dictionary convolutions, PQKD replaces dense convolutional kernels with a low-rank spatial basis whose channel mixing is driven by a compact, shot-limited photonic feature, while keeping inference fully classical. Across MNIST, Fashion-MNIST and CIFAR-10, PQKD achieves stable knowledge distillation under progressively stronger convolutional compression and exposes a clear compression–accuracy frontier controlled by the basis rank and photonic parameter budget. The observed shot-noise behaviour and the effectiveness of simple feature smoothing further indicate that PQKD remains usable under realistic sampling constraints, providing a practical route to integrate near-term photonic quantum hardware into parameter-efficient learning pipelines.

\section{Methods}

\subsection{Datasets and preprocessing}

\paragraph{MNIST.}
We used the MNIST handwritten digit dataset (28$\times$28 grayscale, 10 classes) as the primary benchmark to quantify compression--accuracy trade-offs under controlled conditions. Images were converted to tensors and rescaled to $[0,1]$ without additional augmentation. Unless stated otherwise, we constructed fixed splits of \texttt{train/val} = \texttt{45{,}000/5{,}000} by randomly permuting the canonical MNIST training and test sets using a fixed seed (see \emph{Reproducibility}). Data loaders used batch size 64 with shuffling enabled for training only.

\paragraph{Fashion-MNIST.}
To evaluate PQKD beyond digit recognition and to test robustness on more structured visual patterns, we additionally used the Fashion-MNIST dataset, which consists of 28$\times$28 grayscale images from 10 clothing categories. Preprocessing followed the same protocol as MNIST: images were converted to tensors and rescaled to $[0,1]$ without data augmentation. We constructed fixed \texttt{train/val} splits of \texttt{45{,}000/5{,}000} using a fixed random seed, and used identical data-loading settings. Compared to MNIST, Fashion-MNIST exhibits higher intra-class variability and weaker local stroke structure, providing a more challenging test for structured convolutional compression.

\paragraph{CIFAR-10.}
To assess scalability to natural images with colour channels and richer spatial statistics, we further evaluated PQKD on the CIFAR-10 dataset, comprising 32$\times$32 RGB images across 10 object classes. Images were converted to tensors and normalised channel-wise using the standard CIFAR-10 mean and standard deviation. No data augmentation was applied, in order to isolate the effect of architectural compression. We constructed fixed \texttt{train/val} splits of \texttt{45{,}000/5{,}000} using a fixed seed. For CIFAR-10 experiments, convolutional layers were adjusted to accept three input channels, while all compression mechanisms, photonic feature mappings, and training protocols were otherwise kept identical to the MNIST and Fashion-MNIST settings.

\paragraph{Synthetic dataset.}
To decouple architectural effects from dataset-specific structure and to stress-test robustness, we additionally created a synthetic 10-class image dataset of size-matched 28$\times$28 grayscale images. Each class $c\in\{0,\dots,9\}$ is defined by a small set of class-specific Gaussian ``ink'' templates placed at fixed anchor locations with random jitter:
\begin{enumerate}
\item Start from a zero image $I\in\mathbb{R}^{28\times 28}$.
\item For class $c$, select $m$ anchors $\{(\mu_{j}^{(c)},\nu_{j}^{(c)})\}_{j=1}^m$ (we used $m=2$--$3$), apply jitter $(\Delta\mu,\Delta\nu)\sim\mathcal{N}(0,\sigma_{\mathrm{jit}}^2)$, and add Gaussian blobs
$
I \leftarrow I + \sum_{j=1}^m a_j \exp\!\left(-\frac{(u-\mu_j)^2+(v-\nu_j)^2}{2\sigma_{\mathrm{blob}}^2}\right),
$
with amplitudes $a_j\sim\mathrm{Uniform}(0.8,1.2)$.
\item Add pixel noise $\epsilon\sim\mathcal{N}(0,\sigma_{\mathrm{pix}}^2)$ and clip to $[0,1]$.
\end{enumerate}
We used $\sigma_{\mathrm{blob}}=1.6$, $\sigma_{\mathrm{jit}}=0.8$, and $\sigma_{\mathrm{pix}}=0.05$ unless stated otherwise. The synthetic dataset was class-balanced and split into train/val/test in the same ratios as MNIST. This dataset provides a tunable setting to probe the effect of compression and photonic noise under controlled intra-class variability.

\subsection{Classical models}

\paragraph{Teacher CNN.}
The teacher is a conventional CNN with three convolutional blocks and a global-average-pooling (GAP) classifier head:
\begin{itemize}
\item \textbf{conv1:} $1\rightarrow c_1$, kernel $5\times5$, padding 2, ReLU, $2\times2$ max-pool;
\item \textbf{conv2:} $c_1\rightarrow c_2$, kernel $3\times3$, padding 1, ReLU, $2\times2$ max-pool;
\item \textbf{conv3:} $c_2\rightarrow c_3$, kernel $3\times3$, padding 1, ReLU, dropout ($p=0.25$);
\item \textbf{GAP:} adaptive average pooling to $1\times1$;
\item \textbf{fc:} linear layer $c_3\rightarrow 10$.
\end{itemize}
% We report results for multiple teacher widths (e.g., $(c_1,c_2,c_3)\in\{(32,64,128),(48,96,128),(64,128,128)\}$) to control the teacher capacity and to make compression ratios interpretable.

We report results for multiple teacher widths (e.g., $(c_1,c_2,c_3)\in\{(32,64,128),(48,96,128),(64,128,128)\}$) to control teacher capacity and keep reported compression factors directly comparable across operating points (since the student’s parameterisation is largely fixed by $(R_1,R_2,R_3)$ and $\dim(\theta)$), and to ensure that any performance differences are not an artefact of an over/under-parameterised teacher. With biases enabled and no BatchNorm, the teacher has
\(
P_{\mathrm{teacher}} = 9c_1c_2 + 9c_2c_3 + 26c_1 + c_2 + 11c_3 + 10
\)
trainable parameters. For the widths used in this work:
\(
(c_1,c_2,c_3)=(32,64,128)\Rightarrow P_{\mathrm{teacher}}=94{,}474
\) ; 
\(
(c_1,c_2,c_3)=(48,96,128)\Rightarrow P_{\mathrm{teacher}}=154{,}826
\) ; 
\(
(c_1,c_2,c_3)=(64,128,128)\Rightarrow P_{\mathrm{teacher}}=224{,}394.
\)

\subsection{Quantum model: Photonic Quantum-Enhanced Knowledge Distillation}

\subsubsection{Photonic circuit, sampling, and continuous-variable setting}
PQKD uses a parameterised photonic circuit to generate a compact conditioning vector. We consider a continuous-variable (CV) photonic processor implemented as an integrated optical interferometer with tunable parameters $\theta$ (e.g., phase shifts and beam-splitter angles; optionally including squeezing depending on hardware). For a fixed input state $\rho_{\mathrm{in}}$ over $N$ modes and a parameterised unitary $U(\theta)$, the output state is
\begin{equation}
\rho_{\theta}=U(\theta)\,\rho_{\mathrm{in}}\,U(\theta)^{\dagger}.
\end{equation}

Let $\Omega$ denote a finite (or countable) set of measurement outcomes and let
$\{M_{\omega}\}_{\omega\in\Omega}$ be a POVM on the $N$-mode Hilbert space, i.e.,
$M_{\omega}\succeq 0$ for all $\omega\in\Omega$ and $\sum_{\omega\in\Omega} M_{\omega}=\mathbb{I}$.
Measuring $\rho_\theta$ with this POVM induces the categorical distribution
\begin{equation}
p_{\theta}(\omega)=\Tr\!\big[M_{\omega}\,\rho_{\theta}\big],\qquad \omega\in\Omega,
\label{eq:photonic_distribution-2}
\end{equation}
from which we obtain i.i.d.\ samples $\omega_1,\ldots,\omega_S\sim p_{\theta}$.

In practice, we obtain $S$ i.i.d.\ samples $\omega_1,\ldots,\omega_S\sim p_{\theta}$ (``shots'') per feature evaluation. Experiments reported here use $N=16$ modes with a fixed input pattern and shot budget $S$ held constant within a run (unless varied for noise studies). The primary backend is a photonic sampler consistent with time-bin / interferometric circuit models; when hardware is unavailable, we use a simulator with the same input/output interface.

\subsubsection{Photonic feature construction}
From the $S$ sampled outcomes, we construct a fixed-length feature vector $z(\theta)\in\mathbb{R}^{d}$ with $d=512$:
\begin{enumerate}
\item Convert each sample to a length-$N$ binary vector (thresholded occupancy) $b\in\{0,1\}^{16}$.
\item Split $b$ into two 8-bit halves and map each half to an integer index in $\{0,\dots,255\}$.
\item Form two histograms over $\{0,\dots,255\}$, normalise them to probability vectors, and concatenate to obtain $z\in\mathbb{R}^{512}$.
\item Standardise $z$ (zero mean, unit variance) and apply a fixed scaling factor to control conditioning magnitude.
\end{enumerate}
This marginal-based representation is intentionally low-variance under finite shots and yields a stable conditioning signal without requiring differentiation through the photonic sampler.

\subsubsection{Compressed convolution via photonic-conditioned dictionary mixing}
Each compressed convolutional layer replaces a fully trainable kernel tensor
$W\in\mathbb{R}^{C_{\mathrm{out}}\times C_{\mathrm{in}}\times k\times k}$
by a dictionary model that separates \emph{spatial structure} from \emph{channel mixing}.
Specifically, we introduce $R$ trainable spatial basis filters
$B\in\mathbb{R}^{R\times k\times k}$ and a channel-mixing tensor
$M\in\mathbb{R}^{C_{\mathrm{out}}\times C_{\mathrm{in}}\times R}$ such that, for each output channel $o$,
input channel $i$, and spatial indices $\alpha,\beta\in\{1,\ldots,k\}$,
\begin{equation}
W_{o, i, \alpha, \beta}
=\sum_{r=1}^{R} M_{o,i,r}\,B_{r, \alpha, \beta}.
\label{eq:kernel_recon-2}
\end{equation}
Equivalently, using slice notation one may write
$W_{o,i,:,:}=\sum_{r=1}^{R} M_{o,i,r}\,B_{r,:,:}$, but
Eq.~\eqref{eq:kernel_recon-2} makes the index structure explicit.

Rather than learning $M$ directly, PQKD constrains it through the photonic feature:
\begin{equation}
\mathrm{vec}(M)=A\,z(\theta),
\end{equation}
where $A\in\mathbb{R}^{(C_{\mathrm{out}}C_{\mathrm{in}}R)\times d}$ is fixed at initialisation (Gaussian i.i.d.\ entries with variance $1/d$) and is \emph{not} trained. The student therefore learns only the basis filters $B$ (and biases) in the compressed layers, while the photonic module controls channel mixing via $z(\theta)$.

\paragraph{Layer compression configurations.}
We used three configurations to probe scaling:
\begin{enumerate}
\item \textbf{Compress conv1 only:} replace conv1 by the photonic-conditioned dictionary layer; keep conv2--conv3 dense.
\item \textbf{Compress conv1+conv2:} replace conv1 and conv2; keep conv3 dense.
\item \textbf{Compress all convs:} replace conv1--conv3; keep GAP+fc dense.
\end{enumerate}
Each compressed layer has its own basis rank $R_{\ell}$ (e.g., $R_1,R_2,R_3$), and we swept these ranks to trace compression frontiers.

\subsection{Training objective and optimisation}

\subsubsection{Knowledge distillation loss}
Teacher logits are denoted $t(x)$ and student logits $s(x;w,\theta)$. With temperature $\tau$ and mixing coefficient $\lambda$, we minimise
\begin{equation}
\begin{aligned}
\mathcal{L}_{\mathrm{KD}}(w,\theta)
=&\ \mathbb{E}_{(x,y)}\Big[
\lambda\,\mathrm{CE}\!\big(y,\ \mathrm{softmax}(s(x;w,\theta))\big) \\
&\quad + (1-\lambda)\,\tau^{2}\,
\mathrm{KL}\!\Big(
\mathrm{softmax}\!\big(\tfrac{t(x)}{\tau}\big)\ \Big\Vert\
\mathrm{softmax}\!\big(\tfrac{s(x;w,\theta)}{\tau}\big)
\Big)
\Big].
\end{aligned}
\end{equation}

Unless stated otherwise, we used $\tau=3$ and $\lambda=0.5$.

\subsubsection{Alternating optimisation (student weights and photonic parameters)}
Training proceeds in two nested loops:
\begin{itemize}
\item \textbf{Student update:} with the current photonic feature $z(\theta)$ fixed, update student weights $w$ (basis filters, biases, and classifier head) using Adam on $\mathcal{L}_{\mathrm{KD}}$.
\item \textbf{Photonic update:} update $\theta$ using a sampling-robust, gradient-free method based on a validation proxy of $\mathcal{L}_{\mathrm{KD}}$ while holding $w$ fixed.
\end{itemize}
We used $100$ epochs for the teacher and $100$ epochs for the student. For photonic updates we performed $\texttt{theta\_updates\_per\_epoch}=10$ steps per student epoch. The shot budget $S$ was held fixed (``shots kept'') unless explicitly varied in noise studies.

\subsubsection{Photonic parameter updates (SPSA)}
To avoid backpropagation through shot-noisy sampling, we used simultaneous perturbation stochastic approximation (SPSA). At step $k$, draw a Rademacher perturbation vector $\Delta_k\in\{\pm1\}^{\dim(\theta)}$ and evaluate a validation objective $J(\theta)$ (the KD loss computed on a fixed number of validation batches) at
$\theta_k^{\pm}=\theta_k \pm c\,\Delta_k$.
The gradient estimate is
\begin{equation}
\widehat{g}_k=\frac{J(\theta_k^{+})-J(\theta_k^{-})}{2c}\,\Delta_k,
\end{equation}
and the update is $\theta_{k+1}=\theta_k-a\,\widehat{g}_k$, with clipping to a bounded interval to prevent parameter blow-up. We used constant step sizes $(a,c)$ in the main experiments and verified that results were qualitatively robust to moderate changes in these hyperparameters.

\subsubsection{Feature smoothing and stability}
To reduce shot-to-shot variance, we optionally applied exponential moving average (EMA) smoothing to the photonic feature:
\begin{equation}
\bar z \leftarrow \beta\,\bar z + (1-\beta)\,z(\theta),
\end{equation}
and used $\bar z$ as the conditioning signal for the student forward pass. We used $\beta=0.9$ unless stated otherwise.

\subsection{Noise impact experiments}
To probe robustness under hardware-relevant noise, we conducted three perturbation studies:
\begin{enumerate}
\item \textbf{Finite-shot noise:} vary $S$ (shots per feature evaluation) while keeping all other hyperparameters fixed.
\item \textbf{Feature corruption:} add zero-mean Gaussian noise to the photonic feature, $\tilde z = z + \eta$ with $\eta\sim\mathcal{N}(0,\sigma_z^2 I)$, to emulate readout fluctuations or unmodelled drift in feature statistics.
\item \textbf{Parameter drift:} perturb photonic parameters between epochs, $\theta\leftarrow \theta + \xi$ with $\xi\sim\mathcal{N}(0,\sigma_\theta^2 I)$, to emulate slow hardware drift.
\end{enumerate}
Accuracy and loss were reported as a function of $S$, $\sigma_z$, and $\sigma_\theta$, and compared with and without EMA smoothing to quantify stability improvements.

\subsection{Evaluation metrics and compression accounting}

\paragraph{Accuracy.}
We report top-1 classification accuracy on validation and held-out test sets.

\paragraph{Trainable parameter counts.}
Compression ratios are computed using the number of \emph{trainable} parameters. Fixed matrices $A$ used to generate mixing coefficients are not trainable and are not counted toward adaptation capacity. When reporting end-to-end compression, we include the photonic parameter budget $\dim(\theta)$ as trainable parameters. We separately report convolution-only compression ratios (restricted to the convolutional stack) and full-model compression ratios (including the classifier head).

\subsection{Implementation details and reproducibility}

\paragraph{Software and hardware.}
Classical models were implemented in PyTorch and trained on a CUDA-capable GPU; the PQKD model was also inferred using a CUDA-capable GPU. Photonic sampling was executed via the ORCA simulator backend in CUDA-Q (matching the circuit-level sampling interface used in our pipeline); when available, the same experiment can be run against ORCA Computing's \texttt{ptseries} backend with identical sampling calls and output formatting, enabling drop-in replacement of the simulator by hardware-relevant photonic sampling.

\paragraph{Hyperparameters.}
Unless otherwise stated, we used: batch size 64; Adam learning rate $10^{-3}$ for both teacher and student; dropout $p=0.25$; temperature $\tau=3$; distillation mixing $\lambda=0.5$; $N=16$ photonic modes; feature dimension $d=512$; $\texttt{theta\_updates\_per\_epoch}=10$; and a fixed shot budget $S$ (held constant within an experiment).

\paragraph{Randomness control.}
We fixed random seeds for dataset splits, network initialisation, and stochastic optimisers. For each configuration, we used the same split indices and repeated runs with multiple seeds where noted; summary statistics (mean and standard deviation) are reported when multiple runs were performed.

\section*{Acknowledgment.}
We thank NVIDIA NVAITC and the NVIDIA Quantum team for the helpful discussions and feedback on this research work.
We acknowledge support from the Polish Centre for Supercomputing and Networking (PCSS), Poland, the National Quantum Computing Centre (NQCC), UK and National Center for High-performance Computing (NCHC), Taiwan.
This work was supported by the Engineering and Physical Sciences Research Council (EPSRC) under grant number EP/W032643/1 and Imperial Quantum ICoNYCh Seed Fund. Y.-J. Chang acknowledges support from the National Science and Technology Council (NSTC), Taiwan, under Grants NSTC 114-2112-M-033-010-MY3.

\section*{Data availability.}
MNIST, Fashion-MNIST, and CIFAR-10 are publicly available. The synthetic dataset generator and the exact split indices used in this work can be released alongside the code to enable exact reproduction.

\section*{Code availability.}
The underlying code for this study is not publicly available but may be made available to qualified researchers upon reasonable request to the corresponding author. CUDA-Q is publicly available in NVIDIA’s software library.

\clearpage


\bibliography{sn-bibliography}% common bib file

\clearpage

% --- Supplementary numbering ---
\renewcommand{\thefigure}{S\arabic{figure}}
\renewcommand{\thetable}{S\arabic{table}}
\renewcommand{\thealgorithm}{S\arabic{algorithm}}
\renewcommand{\theequation}{S\arabic{equation}}
\setcounter{figure}{0}
\setcounter{table}{0}
\setcounter{algorithm}{0}
\setcounter{equation}{0} 

\newcommand{\Seqnref}[1]{Eqn~S\ref{#1}}

% Optional: section numbering as S1, S2, ...
\renewcommand{\thesection}{S\arabic{section}}
\renewcommand{\thesubsection}{S\arabic{section}.\arabic{subsection}}
\setcounter{section}{0}

% Optional: page numbering as S1, S2, ...
% \setcounter{page}{1}
% \renewcommand{\thepage}{S\arabic{page}}

% --- Manual supplementary title block (sn-jnl-safe) ---
\begin{center}
{\Large\bfseries Supplementary Information: \\ Photonic Quantum Knowledge Distillation\par}
\vspace{0.8em}
{\normalsize
Kuan-Cheng Chen$^{1,2,7}$,
Shang Yu$^{2,3}$,
Chen-Yu Liu$^{4}$,
Samuel Yen-Chi Chen$^{5}$,
Huan-Hsin Tseng$^{5}$,
Yen Jui Chang$^{6}$,
Wei-Hao Huang$^{7}$,
Esperanza Cuenca Gomez$^{8}$,
Zohim Chandani$^{8}$,
Willian Clements$^{9}$,
Ian Walmsley$^{2,3,10}$,
Kin K.\ Leung$^{1}$\par}
\vspace{0.35em}

{\small\itshape
Kuan-Cheng Chen and Shang Yu contributed equally to this work.\par}
\vspace{0.35em}

{\small
\texttt{kuan-cheng.chen17@imperial.ac.uk}\quad
\texttt{shang.yu@imperial.ac.uk}\quad
\texttt{d11245002@ntu.edu.tw}\quad
\texttt{ycchen1989@ieee.org}\quad
\texttt{htseng@bnl.gov}\quad
\texttt{aceest@cycu.edu.tw}\quad
\texttt{w.huang@j-ij.com}\quad
\texttt{ecuencagomez@nvidia.com}\quad
\texttt{zchandani@nvidia.com}\quad
\texttt{wclements@orcacomputing.com}\quad
\texttt{ian.walmsley@physics.ox.ac.uk}\quad
\texttt{kin.leung@imperial.ac.uk}\par}
\vspace{0.8em}

{\small
$^{1}$ Department of Electrical and Electronic Engineering, Imperial College London, South Kensington, London SW7 2AZ, England, United Kingdom\par
$^{2}$ Imperial Centre for Quantum Engineering, Science and Technology, Imperial College London, South Kensington, London SW7 2AZ, England, United Kingdom\par
$^{3}$ Blackett Laboratory, Department of Physics, Imperial College London, South Kensington, London SW7 2AZ, England, United Kingdom\par
$^{4}$ Graduate Institute of Applied Physics, National Taiwan University, No.\ 1, Sec.\ 4, Roosevelt Rd., Taipei 106319, Taiwan\par
$^{5}$ Computational Science Initiative, Brookhaven National Laboratory, Bldg.\ 725, Room 2-189, P.O.\ Box 5000, Upton, NY 11973-5000, USA\par
$^{6}$ Quantum Information Center, Chung Yuan Christian University, No.\ 200, Zhongbei Rd., Zhongli Dist., Taoyuan City 320314, Taiwan\par
$^{7}$ Jij Inc., Rutherford Appleton Laboratory, Harwell Campus, Didcot OX11 0QX, United Kingdom\par
$^{8}$ NVIDIA Corporation, Santa Clara, CA, USA\par
$^{9}$ ORCA Computing, London, United Kingdom\par
$^{10}$ Department of Physics, University of Oxford, Parks Road, Oxford OX1 3PU, England, United Kingdom\par
}

\end{center}

\section*{Supplementary Contents}
\begin{enumerate}
  \item Supplementary Note 1: Fundamentals 
  \item Supplementary Note 2: Formal definition of PQKD 
  \item Supplementary Note 3: Theory 
  \item Supplementary Note 4: Datasets 
  \item Supplementary Note 5: Classical models 
  \item Supplementary Note 6: Quantum/photonic model 
  \item Supplementary Note 7: Training protocol 
  \item Supplementary Note 8: Experiment setup
\end{enumerate}

\newpage

\section{Supplementary Note 1: Fundamentals}
\label{sec:sn2}

\subsection{Knowledge distillation}
Knowledge distillation \cite{hinton2015distilling,buciluǎ2006model} trains a compact student model to approximate the predictive behaviour of a stronger teacher by using the teacher's \emph{soft} class probabilities as supervision. Let the teacher produce logits $t(x)\in\R^{C}$ and the student produce logits $s(x;w,\theta)\in\R^{C}$ for an input $x$, where $C$ is the number of classes, $w$ denotes the trainable student parameters, and $\theta$ denotes the photonic parameters that condition the student through the feature $z(\theta)$ described in Supplementary Note~\ref{sec:sn3}. For temperature $\tau>0$, define softened distributions
\begin{equation}
p_T^{(\tau)}(x)=\softmax\!\left(\frac{t(x)}{\tau}\right),\qquad
p_S^{(\tau)}(x;w,\theta)=\softmax\!\left(\frac{s(x;w,\theta)}{\tau}\right).
\label{eq:kd_soft}
\end{equation}
Larger $\tau$ increases the entropy of the distributions, revealing ``dark knowledge'' in the teacher's inter-class similarities, which can be particularly informative when the student hypothesis class is strongly constrained by compression.

We optimise the student using a standard distillation objective that interpolates between hard-label supervision and teacher matching:
\begin{equation}
\begin{aligned}
\mathcal{L}_{\mathrm{KD}}(w,\theta)
=&\ \E_{(x,y)}\Big[
\lambda\,\CE\!\big(y,\ p_S^{(1)}(x;w,\theta)\big)
+ (1-\lambda)\,\tau^{2}\,
\KL\!\big(p_T^{(\tau)}(x)\ \Vert\ p_S^{(\tau)}(x;w,\theta)\big)
\Big],
\end{aligned}
\label{eq:kd_loss_supp}
\end{equation}
where $\lambda\in[0,1]$ controls the mixture and $\CE(y,p)$ denotes cross-entropy between the one-hot label $y$ and predicted probability vector $p$. The $\tau^{2}$ factor preserves gradient magnitudes as $\tau$ varies, because the softmax gradients scale approximately as $1/\tau$; without this factor, the distillation signal would diminish at higher temperatures.

The Kullback--Leibler term admits an equivalent and practically useful interpretation. Using $\KL(p\Vert q)=\sum_i p_i\log\frac{p_i}{q_i}$, we can write
\begin{equation}
\KL\!\big(p_T^{(\tau)} \Vert p_S^{(\tau)}\big)
=
-\sum_{c=1}^{C} p_{T,c}^{(\tau)}(x)\,\log p_{S,c}^{(\tau)}(x;w,\theta)
+\sum_{c=1}^{C} p_{T,c}^{(\tau)}(x)\,\log p_{T,c}^{(\tau)}(x),
\label{eq:kd_kl_ce}
\end{equation}
that is,
\begin{equation}
\KL\!\big(p_T^{(\tau)} \Vert p_S^{(\tau)}\big)
=
\CE\!\big(p_T^{(\tau)},\,p_S^{(\tau)}\big)
-\mathrm{H}\!\big(p_T^{(\tau)}\big),
\label{eq:kd_ce_entropy}
\end{equation}
where $\mathrm{H}(p_T^{(\tau)})$ is the Shannon entropy of the teacher distribution and is independent of the student parameters. Consequently, minimising the KL term is equivalent to minimising cross-entropy against the teacher soft targets\cite{hinton2015distilling,cover1999elements}. This encourages the student to reproduce the teacher's \emph{relative class structure} (including non-argmax probabilities), which provides a denser training signal than one-hot labels and empirically improves optimisation and generalisation in compressed regimes.

Finally, in the PQKD setting the distillation signal plays an additional structural role: because the student parameters are restricted to a low-dimensional manifold induced by the photonic-conditioned convolution parameterisation, the teacher soft targets guide the optimisation toward solutions that best approximate the teacher \emph{within this constrained model class}. In this sense, distillation acts as the mechanism that selects, via $(w,\theta)$, a high-performing point on the compression--accuracy frontier reported in the Sec.~\ref{sec:exp1} and Sec.~\ref{sec:exp2}.

\subsection{Model compression and structured parameterisation}

Modern deep networks are frequently over-parameterised relative to the complexity of the target task, yet the resulting parameter budgets can be prohibitive in settings where \emph{training-time} updates, deployment constraints, or communication overhead dominate\cite{zhang2016understanding}. In particular, for edge inference\cite{wang2025low,wang2019adaptive}, federated learning\cite{mcmahan2017communication}, continual learning\cite{parisi2019continual}, and hardware-in-the-loop optimisation\cite{wright2022deep}, the cost of maintaining and updating large weight tensors can exceed the cost of forward computation. Compression methods therefore aim to reduce resource requirements while preserving accuracy, and can be broadly categorised according to the resource being optimised.

We distinguish three complementary notions of compression:
\begin{itemize}
\item \textbf{Trainable-parameter compression:} reducing the number of parameters that must be learned or updated to reach a given performance level. This directly impacts optimisation complexity, storage of trainable state (including optimiser moments), and the communication cost of gradient/parameter synchronisation\cite{wang2025parameter,zheng2019communication}.
\item \textbf{Inference compute reduction:} reducing the floating-point operations (FLOPs), latency, or energy consumed during inference. This is commonly targeted by pruning, low-rank factorisation, or efficient architectural blocks\cite{menghani2023efficient,liu2025survey}.
\item \textbf{Bandwidth/memory reduction:} reducing the cost of storing, transmitting, or streaming model weights and activations, which can be a dominant bottleneck on modern accelerators due to limited memory bandwidth\cite{gholami2024ai,kao2023flat}.
\end{itemize}
These notions are related but not equivalent: for example, a method can sharply reduce trainable parameters while leaving inference FLOPs largely unchanged, or vice versa. In this work, we focus on \emph{trainable-parameter compression} because PQKD is designed as a structured mechanism that restricts the effective degrees of freedom of convolutional layers while allowing the resulting compressed hypothesis class to be tuned by a compact photonic control signal.

Our approach follows the principle of \emph{structured parameterisation}: rather than zeroing or quantising weights post hoc, we explicitly constrain the weight tensors to lie in a low-dimensional family that is efficient to store and update. Convolutional layers are particularly suitable for this strategy because their parameters factor naturally into spatial and channel dimensions. By replacing dense kernels with a dictionary of spatial basis filters and a low-dimensional mixing mechanism (Supplementary Note~\ref{sec:sn3}), PQKD reduces the number of trainable degrees of freedom in the convolutional stack. Knowledge distillation then supplies a dense supervisory signal that mitigates the accuracy loss typically associated with strong constraints, enabling us to trace a controlled compression--accuracy frontier.

\subsection{Continuous-variable photonic quantum computing}
\label{sec:supp_cv_photonics}

Continuous-variable (CV) photonic quantum computing provides a hardware platform in which information is encoded in bosonic modes of the electromagnetic field and processed through optical transformations\cite{braunstein2005quantum,weedbrook2012gaussian}. In integrated implementations, the dominant programmable components are reconfigurable interferometric networks composed of beam splitters and phase shifters, optionally augmented by Gaussian operations such as squeezing (when available)\cite{wang2020integrated,carolan2015universal}. From a learning-systems perspective, a defining characteristic is that readout is naturally \emph{sampling-based}: photon-number-resolving (PNR) detection\cite{deshpande2022quantum} or threshold detection\cite{quesada2018gaussian,hadfield2009single} produces stochastic outcomes whose empirical statistics converge to an underlying measurement distribution as the shot budget $S$ increases.

Photonic quantum knowledge distillation (PQKD) is designed to interface directly with this CV photonic sampling model. Rather than requiring analytic gradients through the photonic circuit---which can be challenging under realistic noise, imperfect calibration, or restricted access to device internals---PQKD treats the photonic processor as a controllable stochastic feature generator. Circuit parameters $\theta$ modulate the output distribution, and a fixed classical feature map converts finite-shot samples into a compact conditioning vector that controls a structured parameterisation of the student network. This separation respects near-term photonic hardware constraints: the photonic device is used in the regime it naturally supports (shot-limited sampling from a programmable optical circuit), while optimisation of the coupled classical--photonic system proceeds with sampling-robust, gradient-free updates for $\theta$ and standard gradient-based updates for the classical student parameters.

A CV photonic circuit can be described as a sequence of elementary gate operations acting on one or more bosonic modes. Each mode is associated with annihilation and creation operators $(\hat a,\hat a^\dagger)$, number operator $\hat n=\hat a^\dagger \hat a$, and quadratures $(\hat x,\hat p)$\cite{braunstein2005quantum}. In practice, the gate set implemented in integrated photonics is dominated by \emph{Gaussian} operations---those generated by Hamiltonians at most quadratic in $(\hat a,\hat a^\dagger)$---including displacements, rotations, squeezing, and multi-mode linear interferometers built from beam splitters and phase shifters\cite{carolan2015universal,reck1994experimental}. Non-Gaussian gates (generated by higher-than-quadratic Hamiltonians), such as Kerr interactions, are more challenging experimentally but are conceptually important because Gaussian operations alone are not universal for quantum computation over CV systems\cite{lloyd1999quantum,menicucci2006universal}. In the experiments considered here, the photonic module is parameterised through a programmable interferometer (a product of beam splitters and rotations) and is accessed through sampling-based measurement; therefore, this gate-level description provides a direct bridge between physical programmability and the algorithmic interface\cite{quesada2018gaussian,deng2023gaussian}.

\paragraph{Gaussian single- and two-mode gates.}
The elementary Gaussian gates used to construct most CV photonic processors include\cite{weedbrook2012gaussian}:
\begin{itemize}
\item \textbf{Displacement.} The displacement gate shifts the phase-space mean of a mode and can be written as
\begin{equation}
\hat D(\alpha)=\exp\!\big(\alpha \hat a^\dagger-\alpha^{\ast}\hat a\big),
\qquad \alpha\in\mathbb{C},
\end{equation}
or equivalently parameterised as $\alpha=a e^{i\theta}$ with amplitude $a\ge 0$ and phase $\theta\in[0,2\pi)$.

\item \textbf{Squeezing.} The squeezing gate changes the variances of conjugate quadratures and generates nonclassical states:
\begin{equation}
\hat S(s)=\exp\!\Big(\tfrac{1}{2}(s^{\ast}\hat a^{2}-s\,\hat a^{\dagger 2})\Big),
\qquad s\in\mathbb{C},
\end{equation}
often written as $s=r e^{i\phi}$ with squeezing strength $r\ge 0$ and phase $\phi\in[0,2\pi)$.

\item \textbf{Rotation (phase shift).} The rotation gate corresponds to an optical phase shift:
\begin{equation}
\hat R(\theta)=\exp(-i\theta \hat n),
\qquad \theta\in[0,2\pi),
\end{equation}
and implements a rotation in phase space, $(\hat x,\hat p)\mapsto(\hat x\cos\theta+\hat p\sin\theta,\,-\hat x\sin\theta+\hat p\cos\theta)$.

\item \textbf{Beam splitter.} A beam splitter couples two modes through interference:
\begin{equation}
\hat{BS}(\theta,\phi)=\exp\!\Big(\theta\big(e^{i\phi}\hat a_1 \hat a_2^\dagger - e^{-i\phi}\hat a_1^\dagger \hat a_2\big)\Big),
\qquad \theta,\phi\in[0,2\pi),
\end{equation}
and is the fundamental two-mode primitive used to build general multi-mode linear interferometers.
\end{itemize}

\paragraph{A representative non-Gaussian gate.}
A canonical example of a non-Gaussian CV gate is the Kerr interaction\cite{lloyd1999quantum},
\begin{equation}
\hat \Phi(\kappa)=\exp(i\kappa \hat n^2),
\qquad \kappa\in[0,2\pi),
\end{equation}
which is generated by a quartic Hamiltonian and can induce highly nonclassical dynamics. While such nonlinearities are generally more demanding to realise in scalable integrated platforms, they are relevant conceptually because adding a suitable non-Gaussian element to Gaussian operations enables universal CV quantum computing. In the present work, we do not rely on explicit Kerr nonlinearities; instead, we exploit sampling-based measurement and classical post-processing as the operational interface to the photonic module.

\paragraph{Interferometers as products of beam splitters and rotations.}
A key workhorse in CV photonic hardware is the $N$-mode linear interferometer, which implements a unitary transformation on the mode operators. In the lossless case, this transformation can be compiled into a sequence of beam splitters and single-mode rotations (phase shifts). Thus, a programmable interferometer can be represented abstractly as
\begin{equation}
\hat U(\boldsymbol{\vartheta})
=\prod_{\ell=1}^{L}\hat{BS}(\theta_\ell,\phi_\ell)\prod_{m=1}^{M}\hat R(\varphi_m),
\end{equation}
where $\boldsymbol{\vartheta}$ collects all tunable angles. This is precisely the type of parameterisation that underlies many integrated meshes \cite{reck1994experimental,clements2016optimal} and loop-based time-bin architectures\cite{motes2014scalable}; in our implementation, this structure manifests as a constraint on the number of beam-splitter parameters per ``tiling'' (Supplementary Note~\ref{sec:sn7}).

\begin{table}[t]
\centering
\caption{\textbf{Common CV gate operations in photonic circuits.} Gaussian gates are generated by Hamiltonians at most quadratic in $(\hat a,\hat a^\dagger)$; Kerr is a representative non-Gaussian gate. In many programmable photonic processors, $\hat U$ is implemented as a product of beam splitters and rotations forming an $N$-port interferometer.}
\label{tab:cv_gates}
\begin{tabular}{@{}llllll@{}}
\toprule
\textbf{Name} & \textbf{Gate} & \textbf{Unitary} & \textbf{Parameters} & \textbf{Gaussian} & \textbf{Modes}\\
\midrule
Displacement & $\hat D(\alpha)$
& $\exp(\alpha \hat a^\dagger-\alpha^\ast \hat a)$
& $\alpha\in\mathbb{C}$ (or $a,\theta$)
& \checkmark & 1\\
Squeezing & $\hat S(s)$
& $\exp\!\big(\tfrac12(s^\ast \hat a^2-s \hat a^{\dagger 2})\big)$
& $s\in\mathbb{C}$ (or $r,\phi$)
& \checkmark & 1\\
Rotation & $\hat R(\theta)$
& $\exp(-i\theta \hat n)$
& $\theta\in[0,2\pi)$
& \checkmark & 1\\
Beam splitter & $\hat{BS}(\theta,\phi)$
& $\exp(\theta(e^{i\phi}\hat a_1 \hat a_2^\dagger-e^{-i\phi}\hat a_1^\dagger \hat a_2))$
& $\theta,\phi\in[0,2\pi)$
& \checkmark & 2\\
Kerr & $\hat \Phi(\kappa)$
& $\exp(i\kappa \hat n^2)$
& $\kappa\in[0,2\pi)$
& $\times$ & 1\\
\bottomrule
\end{tabular}
\end{table}

This gate-level view clarifies how PQKD interfaces with CV photonics. The photonic module is parameterised by tunable interferometric angles (implemented physically through phase shifters and couplers) and is queried through repeated measurement. Rather than requiring explicit analytic gradients through the gate sequence, PQKD uses finite-shot samples to construct a stable feature vector $z(\theta)$, which conditions a structured convolutional parameterisation in the student network. This design aligns the algorithm with the dominant near-term CV photonic capabilities: programmable Gaussian interferometers and sampling-based detection.

% =========================================================
\section{Supplementary Note 2: Definition of Photonic Quantum Knowledge Distillation}
\label{sec:sn3}
\subsection{Photonic quantum feature map}
\label{sec:supp_feature_map}

The role of the photonic module in PQKD is to provide a compact, hardware-aligned \emph{conditioning signal} for the compressed student network\cite{cimini2024variational}. Operationally, the photonic processor is accessed through repeated measurement: for a fixed input preparation and a programmable optical circuit, we observe finite-shot samples from an output distribution whose statistics depend on a low-dimensional parameter vector $\theta$\cite{cerezo2021variational}. PQKD converts these samples into a deterministic, fixed-length feature vector that can be used by standard deep-learning components without requiring an analytic differentiable model of the photonic device.

Formally, let $\rho_{\mathrm{in}}$ denote a fixed $N$-mode input state and let $U(\theta)$ be a parameterised CV unitary representing the programmable photonic transformation. The output state is
\begin{equation}
\rho_{\theta}=U(\theta)\,\rho_{\mathrm{in}}\,U(\theta)^{\dagger}.
\end{equation}
A measurement described by a positive operator-valued measure (POVM) $\{E_{\omega}\}_{\omega\in\Omega}$ induces a probability distribution over outcomes $\omega\in\Omega$,
\begin{equation}
p_{\theta}(\omega)=\Tr\!\big(E_{\omega}\rho_{\theta}\big),
\end{equation}
which captures all device-level effects relevant to the algorithmic interface. In practice, the distribution is not accessed directly; instead, we obtain $S$ independent shots $\omega_1,\ldots,\omega_S\sim p_{\theta}$ and construct an empirical distribution (histogram) $\widehat{p}_{\theta}$. The photonic quantum feature map is then defined as
\begin{equation}
z(\theta)=\Phi(\widehat{p}_{\theta})\in\R^{d},
\end{equation}
where $\Phi$ is a fixed classical post-processing rule that (i) aggregates shot-limited outcomes into a stable representation, (ii) produces a constant-dimensional vector compatible with neural-network layers, and (iii) is inexpensive relative to the downstream classical training.

This design makes explicit that the photonic contribution enters PQKD through \emph{distributional statistics} rather than single-shot values: $z(\theta)$ summarises the circuit-dependent measurement behaviour and serves as a low-dimensional control signal for structured parameter generation in the student network (Supplementary Note~\ref{sec:sn3}). The shot budget $S$ sets the statistical fidelity of $\widehat{p}_{\theta}$ and therefore governs the variance of $z(\theta)$; this dependence motivates the noise and stability analyses reported in Sec.~\ref{sec:exp3} and the use of variance-reduction strategies such as feature smoothing (Supplementary Note~\ref{sec:sn8}).

In the main experiments, we use $N=16$ modes and construct a $d=512$ dimensional feature vector by concatenating two $256$-bin marginal histograms derived from 8-bit partitions of thresholded detection outcomes (Supplementary Note~\ref{sec:sn7}). This choice balances representational richness with robustness under finite shots, and yields a fixed interface that remains unchanged across compression configurations (compressing one, two, or all convolutional layers).
\subsection{Photonic-conditioned dictionary convolution}
\label{sec:supp_pcdconv}

PQKD compresses convolutional layers by replacing a fully trainable kernel tensor with a \emph{structured, low-dimensional parameterisation} whose effective degrees of freedom are controlled by the photonic feature vector $z(\theta)$. This section provides a complete mathematical specification of the parameterisation, its implementation-friendly tensor shapes, and its trainable-parameter accounting.

\paragraph{Standard convolution.}
Consider an input activation map $x\in\R^{C_{\mathrm{in}}\times H\times W}$ and an output activation map $h\in\R^{C_{\mathrm{out}}\times H'\times W'}$. A standard 2D convolution layer with kernel size $k\times k$ and padding $p_{\mathrm{pad}}$ is parameterised by a weight tensor
\begin{equation}
W\in\R^{C_{\mathrm{out}}\times C_{\mathrm{in}}\times k\times k},
\end{equation}
and optional bias $b\in\R^{C_{\mathrm{out}}}$. The output is
\begin{equation}
h_o = \sum_{i=1}^{C_{\mathrm{in}}} W_{o,i} * x_i + b_o,
\qquad o=1,\dots,C_{\mathrm{out}}.
\label{eq:std_conv_def}
\end{equation}
where $*$ denotes spatial convolution and $W_{o,i}\in\R^{k\times k}$ is the kernel slice coupling input channel $i$ to output channel $o$.

\paragraph{Dictionary factorisation of kernels.}
PQKD replaces the dense kernel $W$ by a dictionary (basis) of $R$ spatial filters and a set of channel mixing coefficients. Specifically, introduce trainable basis filters
\begin{equation}
B\in\R^{R\times k\times k},
\end{equation}
and mixing coefficients
\begin{equation}
M\in\R^{C_{\mathrm{out}}\times C_{\mathrm{in}}\times R},
\end{equation}
such that each channel-to-channel kernel is expressed as a linear combination of the shared basis:
\begin{equation}
W_{o,i,:,:}=\sum_{r=1}^{R} M_{o,i,r}\,B_{r,:,:}.
\label{eq:kernel_recon_supp_full}
\end{equation}
Equation~\eqref{eq:kernel_recon_supp_full} is a \emph{structured} parameterisation: spatial degrees of freedom are concentrated in the basis $B$, while channel coupling is handled by $M$. If $R\ll C_{\mathrm{out}}C_{\mathrm{in}}$, this can drastically reduce the number of trainable parameters \emph{provided that} $M$ itself is not fully trainable.

\paragraph{Photonic-conditioned mixing (core PQKD mechanism).}
The key idea in PQKD is to avoid training the full mixing tensor $M$. Instead, $M$ is generated deterministically from the photonic feature $z(\theta)\in\R^{d}$ via a fixed linear map:
\begin{equation}
\mathrm{vec}(M)=A\,z(\theta),
\qquad
A\in\R^{(C_{\mathrm{out}}C_{\mathrm{in}}R)\times d}.
\label{eq:mixing_linear_supp_full}
\end{equation}
Here $\mathrm{vec}(\cdot)$ denotes vectorisation that stacks all entries of $M$ into a column vector. The matrix $A$ is \emph{fixed} (sampled once using a prescribed random seed and then held constant) and is therefore excluded from trainable parameter counts. Under this construction, the channel mixing coefficients $M$ are not free parameters; they vary only through the $d$-dimensional photonic control signal $z(\theta)$.

For implementation, one typically reshapes $A z(\theta)$ back into the tensor form
\begin{equation}
M(z)=\mathrm{reshape}\big(Az,\ [C_{\mathrm{out}}, C_{\mathrm{in}}, R]\big).
\label{eq:mixing_reshape}
\end{equation}
Combining Eq.~\eqref{eq:kernel_recon_supp_full} with Eq.~\eqref{eq:mixing_reshape} yields an explicit mapping from $z(\theta)$ and the basis $B$ to the full kernel:
\begin{equation}
W(z,B)_{o,i,:,:}=\sum_{r=1}^{R} \Big(\mathrm{reshape}(Az,\ [C_{\mathrm{out}}, C_{\mathrm{in}}, R])\Big)_{o,i,r}\,B_{r,:,:}.
\label{eq:kernel_from_zB}
\end{equation}
In other words, \emph{all} channel mixing across the layer is confined to the $d$-dimensional subspace $\{Az: z\in\R^{d}\}$.

\paragraph{Equivalent matrix formulation.}
Let $W_{(k)}\in\R^{(C_{\mathrm{out}}C_{\mathrm{in}})\times k^2}$ denote the kernel tensor flattened over spatial dimensions, and let $B_{(k)}\in\R^{R\times k^2}$ denote the basis filters flattened similarly. Define $M_{(c)}\in\R^{(C_{\mathrm{out}}C_{\mathrm{in}})\times R}$ as the mixing matrix obtained by flattening the first two indices of $M$. Then Eq.~\eqref{eq:kernel_recon_supp_full} is equivalent to
\begin{equation}
W_{(k)} = M_{(c)}\,B_{(k)}.
\label{eq:matrix_fact}
\end{equation}
Equation~\eqref{eq:mixing_linear_supp_full} specifies that $M_{(c)}$ lies in a $d$-dimensional linear subspace parameterised by $z(\theta)$. This makes explicit how PQKD imposes a \emph{low-dimensional manifold} constraint on the kernel space.

\paragraph{Forward computation.}
Given $z(\theta)$ and basis filters $B$, the layer computes:
\begin{enumerate}
\item Generate $M \leftarrow \mathrm{reshape}(A z(\theta), [C_{\mathrm{out}},C_{\mathrm{in}},R])$.
\item Form the effective kernel using Eq.~\eqref{eq:kernel_recon_supp_full}.
\item Apply convolution as in Eq.~\eqref{eq:std_conv_def}.
\end{enumerate}
In practice, step (2) can be implemented efficiently via tensor contractions. One convenient expression is
\begin{equation}
W(z,B)=\mathrm{einsum}\big(\texttt{oir,rxy}\rightarrow\texttt{oixy},\ M(z),\ B\big),
\label{eq:einsum_impl}
\end{equation}
followed by a standard convolution call $\mathrm{conv2d}(x,W,b,\mathrm{padding}=p_{\mathrm{pad}})$.

\paragraph{Trainable-parameter accounting and compression ratio.}
For a single compressed convolution layer, the trainable parameters are:
\begin{equation}
\#\mathrm{params}_{\mathrm{train}} = \underbrace{Rk^2}_{\text{basis }B} + \underbrace{C_{\mathrm{out}}}_{\text{bias}},
\label{eq:trainable_count_layer}
\end{equation}
plus the photonic parameter budget $\dim(\theta)$ if $\theta$ is included in the overall student budget. In contrast, a dense convolution has
\begin{equation}
\#\mathrm{params}_{\mathrm{dense}} = C_{\mathrm{out}}C_{\mathrm{in}}k^2 + C_{\mathrm{out}}.
\label{eq:dense_count_layer}
\end{equation}
Ignoring bias terms for intuition, the per-layer trainable-parameter compression factor is approximately
\begin{equation}
\mathrm{CR}_{\mathrm{layer}}
\approx
\frac{C_{\mathrm{out}}C_{\mathrm{in}}k^2}{Rk^2}
=
\frac{C_{\mathrm{out}}C_{\mathrm{in}}}{R},
\qquad (R\ll C_{\mathrm{out}}C_{\mathrm{in}}).
\label{eq:cr_layer_simple}
\end{equation}
When multiple layers share the same photonic feature $z(\theta)$ (as in our implementation), $\theta$ becomes a \emph{shared} overhead that does not scale with layer width, which is advantageous when compressing multiple convolutional layers simultaneously.

\paragraph{Interpretation and relation to hypernetworks.}
Equation~\eqref{eq:mixing_linear_supp_full} can be viewed as a constrained hypernetwork that generates convolutional mixing coefficients from a compact input $z(\theta)$. The constraint that the generator is linear and fixed (through $A$) intentionally shifts learning capacity away from the generator and into (i) the trainable basis $S$ and (ii) the photonic parameters $\theta$ that control $z(\theta)$ through the sampling-based feature map (Supplementary Note~\ref{sec:supp_feature_map}). This design choice reduces trainable degrees of freedom, improves reproducibility, and makes the photonic contribution explicit and measurable through the dependence on $\theta$ and the shot budget $S$.

\paragraph{Extension to multiple compressed layers.}
For a network with multiple compressed convolutional layers $\ell\in\mathcal{L}$, each layer has its own $(C^{(\ell)}_{\mathrm{in}},C^{(\ell)}_{\mathrm{out}},k_\ell,R_\ell)$, basis $S^{(\ell)}$, and fixed projection $A^{(\ell)}$, while sharing the same $z(\theta)$:
\begin{equation}
\mathrm{vec}(M^{(\ell)}) = A^{(\ell)} z(\theta),\qquad
W^{(\ell)}_{o,i,:,:}=\sum_{r=1}^{R_\ell} M^{(\ell)}_{o,i,r}\,B^{(\ell)}_{r,:,:}.
\label{eq:multi_layer}
\end{equation}
This yields a coherent mechanism to study ``compress conv1'', ``compress conv1+conv2'', and ``compress all convs'' settings under a unified control signal and parameter-counting convention.
\subsection{Compression configurations}
\label{sec:supp_compression_configs}

To quantify how PQKD scales with compression strength and network depth, we consider three structured compression regimes that progressively replace larger fractions of the convolutional stack by photonic-conditioned dictionary convolutions (Supplementary Note~\ref{sec:supp_pcdconv}). Throughout, the teacher network uses standard dense convolutions, while the student network matches the teacher macro-architecture (same number of blocks, pooling schedule, and head) but substitutes selected convolution layers by the PQKD parameterisation. This design isolates the effect of structured parameterisation and the photonic conditioning signal from confounding architectural changes.

\paragraph{Regime I: compress \texttt{conv1} only.}
In this regime, only the first convolution layer is replaced by the photonic-conditioned dictionary convolution:
\begin{equation}
\mathrm{vec}(M^{(1)}) = A^{(1)} z(\theta),\qquad
W^{(1)}_{o,i,:,:}=\sum_{r=1}^{R_{1}} M^{(1)}_{o,i,r}\,S^{(1)}_{r,:,:}.
\end{equation}
All subsequent convolutions (e.g., \texttt{conv2} and \texttt{conv3}) remain dense and fully trainable. This setting represents a conservative compression strategy that primarily reduces parameters near the input while leaving downstream representational capacity largely intact. It provides a baseline to test whether PQKD can compress early feature extraction without destabilising training.

\paragraph{Regime II: compress \texttt{conv1+conv2}.}
Here, both the first and second convolution layers are replaced by PQKD layers with (potentially distinct) basis ranks $R_1$ and $R_2$:
\begin{equation}
\mathrm{vec}(M^{(\ell)}) = A^{(\ell)} z(\theta),\qquad
W^{(\ell)}_{o,i,:,:}=\sum_{r=1}^{R_{\ell}} M^{(\ell)}_{o,i,r}\,S^{(\ell)}_{r,:,:},
\qquad \ell\in\{1,2\}.
\end{equation}
The third convolution remains dense. This regime substantially increases the fraction of trainable convolution parameters controlled by $z(\theta)$ while still retaining one fully trainable convolution block to absorb residual modelling error. Empirically, this intermediate regime often yields a favourable trade-off between compression and accuracy because it compresses a large portion of the convolutional stack while maintaining a flexible downstream adaptation layer.

\paragraph{Regime III: compress all convolution layers (\texttt{conv1--conv3}).}
In the most aggressive setting, \emph{all} convolution layers are replaced by photonic-conditioned dictionary convolutions:
\begin{equation}
\mathrm{vec}(M^{(\ell)}) = A^{(\ell)} z(\theta),\qquad
W^{(\ell)}_{o,i,:,:}=\sum_{r=1}^{R_{\ell}} M^{(\ell)}_{o,i,r}\,S^{(\ell)}_{r,:,:},
\qquad \ell\in\{1,2,3\}.
\label{eq:all_conv_compress}
\end{equation}
This regime tests large-scale structured compression, in which the convolutional feature extractor is entirely restricted to the PQKD manifold defined by $\{A^{(\ell)}z(\theta)\}$ and the trainable bases $\{S^{(\ell)}\}$. Since the same $z(\theta)$ conditions every compressed layer, the photonic parameters $\theta$ act as a \emph{shared global control} over the entire convolutional stack, while each layer retains local spatial expressivity through its basis $S^{(\ell)}$ and rank $R_\ell$.

\paragraph{Rank selection and interpretability.}
Each compressed layer $\ell$ has an associated basis rank $R_\ell$, which controls the number of spatial basis filters in that layer. Intuitively, larger $R_\ell$ increases the spatial expressivity of the reconstructed kernels (Eq.~\eqref{eq:kernel_recon_supp_full}) at the cost of more trainable parameters ($R_\ell k_\ell^2$). In contrast, the channel mixing degrees of freedom remain confined to the $d$-dimensional subspace induced by $A^{(\ell)}z(\theta)$ (Eq.~\eqref{eq:mixing_linear_supp_full}). Our sweep studies therefore treat $\{R_\ell\}$ and $\dim(\theta)$ as primary compression knobs.

\paragraph{Parameter accounting across regimes.}
For a student model with compressed layers $\mathcal{L}$, the total trainable parameters are
\begin{equation}
\#\mathrm{params}_{\mathrm{student}}
=
\sum_{\ell\in\mathcal{L}}\big(R_\ell k_\ell^2 + C^{(\ell)}_{\mathrm{out}}\big)
+
\sum_{\ell\notin\mathcal{L}}\big(C^{(\ell)}_{\mathrm{out}} C^{(\ell)}_{\mathrm{in}} k_\ell^2 + C^{(\ell)}_{\mathrm{out}}\big)
+
\#\mathrm{params}_{\mathrm{head}}
+
\dim(\theta),
\label{eq:student_param_total}
\end{equation}
where $\#\mathrm{params}_{\mathrm{head}}$ denotes the parameters of the non-convolutional head (e.g., global-average-pooling classifier). We report both an \emph{overall} compression ratio (teacher parameters divided by student parameters, including $\dim(\theta)$) and a \emph{conv-only} compression ratio restricted to the convolutional layers, to disentangle compression achieved in the feature extractor from contributions of the classifier head.

\paragraph{Practical implementation.}
In all regimes, each compressed layer uses its own fixed projection matrix $A^{(\ell)}$ (sampled once and held constant) and its own trainable basis $S^{(\ell)}$, while sharing the same photonic feature vector $z(\theta)$ across layers. This choice keeps the student trainable parameter count low and makes the effect of $\theta$ and the shot budget $S$ directly comparable across compression regimes.

% =========================================================
\section{Supplementary Note 3: Theory---expressivity, approximation, and stability}
\label{sec:sn4}
\subsection{Trainable-parameter accounting}
\label{sec:supp_param_accounting}

To report compression ratios consistently, we distinguish \emph{trainable} parameters (those updated by learning) from \emph{fixed} parameters (constants sampled once and held throughout training). This distinction is essential for PQKD because the channel-mixing generator uses a fixed projection matrix $A$ (Eq.~\eqref{eq:mixing_linear_supp_full}), which can be large in memory but does not contribute to the number of learnable degrees of freedom.

\paragraph{Dense convolution (baseline).}
A standard 2D convolution with kernel size $k\times k$, $C_{\mathrm{in}}$ input channels and $C_{\mathrm{out}}$ output channels has trainable parameters
\begin{equation}
\#\mathrm{params}_{\mathrm{dense}}
=
C_{\mathrm{out}}C_{\mathrm{in}}k^2 + C_{\mathrm{out}},
\label{eq:dense_param_count}
\end{equation}
where the first term is the weight tensor $W\in\R^{C_{\mathrm{out}}\times C_{\mathrm{in}}\times k\times k}$ and the second term is the bias vector $b\in\R^{C_{\mathrm{out}}}$.

\paragraph{PQKD compressed convolution.}
In PQKD, a compressed convolution layer uses trainable spatial basis filters
$B\in\R^{R\times k\times k}$ and a bias $b\in\R^{C_{\mathrm{out}}}$, while the channel mixing coefficients are generated deterministically as $\mathrm{vec}(M)=A z(\theta)$ with fixed $A$ and photonic feature $z(\theta)$.
Therefore, the trainable parameters \emph{per compressed layer} are
\begin{equation}
\#\mathrm{params}_{\mathrm{PQKD\ layer}}
=
\underbrace{Rk^2}_{\text{basis }B}
+
\underbrace{C_{\mathrm{out}}}_{\text{bias}},
\label{eq:pqkd_layer_param_count}
\end{equation}
and this count is independent of $C_{\mathrm{in}}$ because channel mixing is not learned directly.

\paragraph{Global photonic parameter budget.}
The photonic parameters $\theta\in\R^{\dim(\theta)}$ are shared across all compressed layers through the feature vector $z(\theta)$. When reporting \emph{overall} student trainable parameters, we include this shared budget once:
\begin{equation}
\#\mathrm{params}_{\mathrm{student}}
=
\sum_{\ell\in\mathcal{L}}\big(R_\ell k_\ell^2 + C_{\mathrm{out}}^{(\ell)}\big)
+
\sum_{\ell\notin\mathcal{L}}\big(C_{\mathrm{out}}^{(\ell)}C_{\mathrm{in}}^{(\ell)}k_\ell^2 + C_{\mathrm{out}}^{(\ell)}\big)
+
\#\mathrm{params}_{\mathrm{head}}
+
\dim(\theta),
\label{eq:student_params_total_repeat}
\end{equation}
where $\mathcal{L}$ denotes the set of compressed convolution layers and $\#\mathrm{params}_{\mathrm{head}}$ denotes any non-convolutional parameters (e.g., the GAP classifier head).

\paragraph{Per-layer compression factor.}
A useful diagnostic is the per-layer trainable-parameter compression factor, obtained by comparing Eq.~\eqref{eq:dense_param_count} and Eq.~\eqref{eq:pqkd_layer_param_count}. Ignoring the shared $\dim(\theta)$ overhead (which is amortised across layers), we define
\begin{equation}
\mathrm{CR}_{\mathrm{conv}}
=
\frac{\#\mathrm{params}_{\mathrm{dense}}}{\#\mathrm{params}_{\mathrm{PQKD\ layer}}}
\approx
\frac{C_{\mathrm{out}}C_{\mathrm{in}}k^2 + C_{\mathrm{out}}}{Rk^2 + C_{\mathrm{out}}}
=
\frac{C_{\mathrm{out}}(C_{\mathrm{in}}k^2+1)}{Rk^2 + C_{\mathrm{out}}}.
\label{eq:cr_conv_supp}
\end{equation}
When $C_{\mathrm{out}}C_{\mathrm{in}}k^2 \gg C_{\mathrm{out}}$ and $Rk^2 \gg C_{\mathrm{out}}$ (a common regime for moderate-to-large channel counts), this simplifies to the intuitive approximation
\begin{equation}
\mathrm{CR}_{\mathrm{conv}} \approx \frac{C_{\mathrm{out}}C_{\mathrm{in}}}{R}.
\label{eq:cr_simplified}
\end{equation}
This highlights the role of $R$ as the primary knob controlling trainable degrees of freedom in the compressed convolution.

\paragraph{Reporting conventions used in this work.}
We report two complementary ratios:
\begin{itemize}
\item \textbf{Conv-only ratio:} computed by restricting Eq.~\eqref{eq:student_params_total_repeat} to convolution layers only. This isolates compression achieved in the feature extractor.
\item \textbf{Overall ratio (including $\theta$):} computed using Eq.~\eqref{eq:student_params_total_repeat}, including $\dim(\theta)$ once, to reflect the full trainable budget required by PQKD.
\end{itemize}
Fixed projection matrices $\{A^{(\ell)}\}$ are excluded from trainable counts throughout because they are not updated during training; however, we note that they can contribute to memory footprint and can be implemented implicitly (e.g., via structured random features) if memory becomes a constraint.

\paragraph{Mixing subspace induced by a fixed generator.}
For a given layer with mixing tensor $M\in\R^{C_{\mathrm{out}}\times C_{\mathrm{in}}\times R}$, define $m=\mathrm{vec}(M)\in\R^{p}$ where
\begin{equation}
p := C_{\mathrm{out}}C_{\mathrm{in}}R.
\end{equation}
PQKD generates $m$ as
\begin{equation}
m = A z,\qquad A\in\R^{p\times d},\quad z\in\R^{d}.
\label{eq:mAz}
\end{equation}
Therefore, all attainable mixing vectors lie in the column space of $A$:
\begin{equation}
m \in \mathrm{Range}(A) := \{Az : z\in\R^{d}\}.
\end{equation}
If $A$ has full column rank (a typical case when $d\le p$ and $A$ is drawn from a continuous distribution), then $\dim(\mathrm{Range}(A))=d$.

\paragraph{Best achievable approximation to a target mixing.}
Let $M^\star$ denote an arbitrary target mixing tensor (e.g., the mixing that would reproduce a desired dense kernel given a fixed basis). Let $m^\star=\mathrm{vec}(M^\star)\in\R^{p}$. The best approximation within the PQKD subspace is the solution of the least-squares problem
\begin{equation}
z^\star = \arg\min_{z\in\R^{d}} \|m^\star - A z\|_2^2.
\label{eq:lsq_def}
\end{equation}
When $A$ has full column rank, the minimiser is unique and given by
\begin{equation}
z^\star = (A^\top A)^{-1}A^\top m^\star,
\label{eq:lsq_solution}
\end{equation}
and the corresponding attainable mixing is the orthogonal projection of $m^\star$ onto $\mathrm{Range}(A)$:
\begin{equation}
A z^\star = \Pi_{\mathrm{Range}(A)}\, m^\star,
\qquad
\Pi_{\mathrm{Range}(A)} := A(A^\top A)^{-1}A^\top.
\label{eq:projection_supp}
\end{equation}
The residual $\|m^\star - A z^\star\|_2$ quantifies the \emph{subspace approximation error} introduced by restricting $M$ to $\mathrm{Range}(A)$.

\paragraph{Implications for expressivity and the role of $d$ and $R$.}
Two independent design choices control the representational capacity of a compressed convolution:
\begin{itemize}
\item \textbf{Mixing dimension $d$:} Increasing $d$ enlarges $\mathrm{Range}(A)$ and reduces the best-case subspace approximation error in Eq.~\eqref{eq:projection_supp}, enabling richer channel coupling patterns across $(C_{\mathrm{out}},C_{\mathrm{in}})$ for a fixed rank $R$.
\item \textbf{Basis rank $R$:} Increasing $R$ enlarges the dictionary $S\in\R^{R\times k\times k}$ and thus increases \emph{spatial} expressivity in the kernel reconstruction $W=M S$ (Eq.~\eqref{eq:matrix_fact}). Even with perfect mixing, a small $R$ limits the diversity of spatial filters that can be expressed.
\end{itemize}
Consequently, $d$ and $R$ play complementary roles: $d$ controls how flexibly PQKD can represent channel-to-channel mixing coefficients, whereas $R$ controls how flexibly it can represent spatial kernel structure. Our sweep studies therefore treat $(d,\dim(\theta))$ and $\{R_\ell\}$ as primary axes when constructing compression--accuracy frontiers (Sec.~\ref{sec:exp2}).

\paragraph{Connection to the photonic quantum feature map.}
In PQKD, the vector $z$ is not a free optimisation variable but is obtained as $z(\theta)=\Phi(\widehat{p}_\theta)$ from shot-limited samples (Supplementary Note~\ref{sec:supp_feature_map}). Thus, Eq.~\eqref{eq:projection_supp} characterises an \emph{idealised best case}: it describes the closest mixing achievable if $z$ could be chosen arbitrarily in $\R^d$. In practice, the attainable set is further constrained by the photonic map $\theta\mapsto z(\theta)$ and finite-shot variability; these effects are investigated empirically via shot and noise sweeps (Sec.~\ref{sec:exp3}).

\subsection{Approximation within the mixing subspace}
\label{sec:supp_mixing_subspace}
Fix the basis $B$ and consider any target mixing tensor $M^{\star}$. Under Eq.~\eqref{eq:mAz}, the best achievable mixing corresponds to the least-squares projection of $\mathrm{vec}(M^{\star})$ onto the column space of $A$:
\begin{equation}
z^{\star}=\arg\min_{z\in\mathbb{R}^{d}}\ \left\|\mathrm{vec}(M^{\star})-Az\right\|_2^{2},
\quad\Rightarrow\quad
Az^{\star}=\Pi_{\mathrm{Range}(A)}\,\mathrm{vec}(M^{\star}),
\label{eq:projection}
\end{equation}
where $\Pi_{\mathrm{Range}(A)}$ denotes orthogonal projection. This characterises how $d$ and the choice of $A$ control the representational subspace for channel mixing, while $R$ controls the spatial expressivity through the basis filters.

\subsection{Finite-shot noise propagation}
\label{sec:supp_finite_shot}

The photonic feature vector $z(\theta)$ is computed from a finite number of measurement shots. Consequently, the PQKD student does not receive the exact distributional feature $z(\theta)=\Phi(p_\theta)$, but rather an empirical estimate $\widehat{z}(\theta)=\Phi(\widehat{p}_\theta)$ constructed from $S$ i.i.d.\ samples. This section formalises how finite-shot fluctuations propagate through the feature map and into the reconstructed convolution kernels, providing a principled basis for the noise studies reported in Sec.~\ref{sec:exp2}.

\paragraph{Concentration of the empirical measurement distribution.}
Let $\widehat{p}_\theta$ be the empirical histogram over outcomes $\omega\in\Omega$ formed from $S$ independent samples $\omega_1,\dots,\omega_S\sim p_\theta$. For any fixed bin $\omega$, the random variable $\widehat{p}_\theta(\omega)$ is the sample mean of Bernoulli trials with mean $p_\theta(\omega)$. Hoeffding's inequality yields, for any $\epsilon>0$,
\begin{equation}
\mathbb{P}\!\left(\left|\widehat{p}_{\theta}(\omega)-p_{\theta}(\omega)\right|\ge \epsilon\right)
\le 2\exp\!\left(-2S\epsilon^2\right).
\label{eq:hoeffding_bin}
\end{equation}
This makes explicit the $1/\sqrt{S}$ scaling of shot noise at the level of individual histogram entries.

\paragraph{From histogram noise to feature noise.}
The photonic feature is obtained via a deterministic map $\Phi$ applied to the empirical histogram: $\widehat{z}(\theta)=\Phi(\widehat{p}_\theta)$ and $z(\theta)=\Phi(p_\theta)$. Suppose $\Phi$ is Lipschitz with respect to the $\ell_1$ norm on histograms, i.e.,
\begin{equation}
\|\Phi(q)-\Phi(q')\|_2 \le L_\Phi \|q-q'\|_1
\qquad \text{for all histograms } q,q'.
\label{eq:lipschitz_phi}
\end{equation}
This assumption is satisfied by histogram-based feature constructions composed of linear operations, normalisation away from degenerate cases, and bounded affine rescaling; it holds for our marginal-histogram mapping under standard numerical stabilisers (Supplementary Note~\ref{sec:sn7}). Applying Eq.~\eqref{eq:lipschitz_phi} gives
\begin{equation}
\|\widehat{z}(\theta)-z(\theta)\|_2
\le L_{\Phi}\|\widehat{p}_{\theta}-p_{\theta}\|_1.
\label{eq:feature_lipschitz_bound}
\end{equation}
To interpret $\|\widehat{p}_\theta-p_\theta\|_1$, let $K$ denote the effective number of histogram bins used by $\Phi$ (e.g., $K=512$ for two concatenated 256-bin marginals). Standard multinomial concentration bounds imply that
\begin{equation}
\E\!\left[\|\widehat{p}_{\theta}-p_{\theta}\|_1\right]
= \mathcal{O}\!\left(\sqrt{\frac{K}{S}}\right),
\end{equation}
and therefore
\begin{equation}
\E\!\left[\|\widehat{z}(\theta)-z(\theta)\|_2\right]
= \mathcal{O}\!\left(L_\Phi\sqrt{\frac{K}{S}}\right).
\label{eq:feature_concentration_supp}
\end{equation}
Equation~\eqref{eq:feature_concentration_supp} provides a direct link between shot budget $S$, feature dimension $K$, and the magnitude of stochastic feature perturbations seen by the student network.

\paragraph{Propagation to mixing coefficients and convolution kernels.}
In PQKD, the mixing tensor is generated linearly from the feature:
\begin{equation}
\mathrm{vec}(M)=A z,\qquad \mathrm{vec}(\widehat{M})=A \widehat{z},
\end{equation}
so the induced perturbation obeys
\begin{equation}
\|\mathrm{vec}(\widehat{M})-\mathrm{vec}(M)\|_2
\le \|A\|_2\,\|\widehat{z}-z\|_2.
\label{eq:mixing_perturb}
\end{equation}
Next, the reconstructed kernel satisfies $W=M B$ in the flattened formulation (Eq.~\eqref{eq:matrix_fact}). Using submultiplicativity of operator norms, one obtains a Lipschitz bound for the kernel perturbation:
\begin{equation}
\|W(B,\widehat{z})-W(B,z)\|_F
=
\|\widehat{M}B - MB\|_F
\le \|\widehat{M}-M\|_F\,\|B\|_2
\le \|A\|_2\,\|B\|_F\,\|\widehat{z}-z\|_2.
\label{eq:kernel_lipschitz_supp}
\end{equation}
where the final inequality uses $\|B\|_2\le \|S\|_F$ and the relationship between Frobenius and Euclidean norms under vectorisation.

\paragraph{Implications for training stability and noise studies.}
Equations~\eqref{eq:feature_concentration_supp}--\eqref{eq:kernel_lipschitz_supp} show that the effective kernel noise induced by finite shots scales approximately as $\mathcal{O}(\sqrt{K/S})$, amplified by the projection magnitude $\|A\|_2$ and the basis energy $\|B\|_F$. This motivates two practical considerations that we evaluate empirically: (i) increasing $S$ reduces stochasticity in the photonic conditioning signal, and (ii) variance-reduction strategies such as feature smoothing (e.g., EMA in Supplementary Note~\ref{sec:sn8}) can reduce the high-frequency fluctuations of $\widehat{z}(\theta)$, thereby stabilising both the induced kernels and the downstream student optimisation. These predictions are directly tested in the shot and noise impact experiments in the Sec.~\ref {sec:exp3}.

% =========================================================
% \section{Supplementary Note 4: Datasets}
% \label{sec:sn5}

% \subsection{MNIST}
% MNIST consists of $28\times 28$ grayscale images over 10 classes. Images are rescaled to $[0,1]$. Unless otherwise stated, we use fixed splits with \texttt{train/val/test} specified in the main text (and stored indices for exact reproducibility).

% \subsection{CIFAR-10}
% CIFAR-10 consists of $32\times 32$ RGB images over 10 classes. We use standard per-channel normalization (mean/std reported in code) and standard light augmentation for training (random crop with padding and random horizontal flip). Validation and test use deterministic preprocessing only.

% \subsection{Synthetic controlled dataset}
% To isolate architectural effects and tune task difficulty, we additionally use a synthetic 10-class dataset of $28\times28$ grayscale images. Each class is defined by $m$ Gaussian ``ink'' blobs at class-specific anchors with jitter:
% \begin{equation}
% I(u,v)=\sum_{j=1}^{m} a_j
% \exp\!\left(-\frac{(u-\mu_j)^2+(v-\nu_j)^2}{2\sigma_{\mathrm{blob}}^2}\right)+\epsilon,
% \end{equation}
% where $(\mu_j,\nu_j)$ are anchor locations perturbed by Gaussian jitter, amplitudes $a_j$ are sampled from a bounded range, and $\epsilon$ is additive pixel noise clipped to $[0,1]$. Default parameters are provided in code; we use class-balanced splits matching MNIST sizes.

% =========================================================
\section{Supplementary Note 5: Classical models and baselines}
\label{sec:sn6}

This Supplementary Note specifies the classical teacher and student architectures used to evaluate PQKD, together with the baseline models required for fair attribution. Our design principle is to control for confounding factors: the teacher and PQKD student share the same high-level topology (number of convolutional blocks, pooling schedule, and classifier head), and we vary \emph{only} the parameterisation of selected convolutional kernels and the source of the conditioning signal. This ensures that observed performance differences can be attributed to structured compression and photonic conditioning, rather than to unrelated architectural changes.

\subsection{Teacher CNN (global-average-pooling head)}
\label{sec:supp_teacher}

The teacher network is a conventional convolutional classifier with three convolutional blocks and a global-average-pooling (GAP) head\cite{lin2013network}, chosen for (i) strong performance on MNIST/CIFAR-10 without excessive depth, (ii) stable training under standard optimisers, and (iii) compatibility with structured compression of the convolutional stack. Let $C_{\mathrm{in}}$ denote the input channels and let $(c_1,c_2,c_3)$ denote the channel widths.

\paragraph{Architecture.}
The teacher consists of:
\begin{itemize}
\item \textbf{conv1:} $C_{\mathrm{in}}\rightarrow c_1$ with $5\times5$ kernels, padding 2, followed by ReLU and $2\times2$ max-pooling.
\item \textbf{conv2:} $c_1\rightarrow c_2$ with $3\times3$ kernels, padding 1, followed by ReLU and $2\times2$ max-pooling.
\item \textbf{conv3:} $c_2\rightarrow c_3$ with $3\times3$ kernels, padding 1, followed by ReLU and dropout with $p=0.25$.
\item \textbf{GAP:} adaptive average pooling to spatial size $1\times1$.
\item \textbf{Classifier:} a linear layer mapping $c_3\rightarrow C$ with $C=10$ classes.
\end{itemize}

\paragraph{Teacher width sweep.}
To obtain clear compression ratios while maintaining a strong teacher reference, we sweep teacher widths over representative settings such as
\begin{equation}
(c_1,c_2,c_3)\in\{(32,64,128),\ (48,96,128),\ (64,128,128)\},
\end{equation}
and report teacher accuracy and parameter counts for each setting. In all cases, the teacher is trained to convergence (or a fixed epoch budget) using standard cross-entropy on hard labels.

\paragraph{Teacher parameter count (convolutional stack).}
For completeness, the number of trainable parameters in the three convolutional layers is
\begin{equation}
\#\mathrm{params}_{\mathrm{teacher,conv}}
=
\sum_{\ell=1}^{3}\Big(C_{\mathrm{out}}^{(\ell)}C_{\mathrm{in}}^{(\ell)}k_\ell^2 + C_{\mathrm{out}}^{(\ell)}\Big),
\end{equation}
with $(k_1,k_2,k_3)=(5,3,3)$ and $(C_{\mathrm{out}}^{(1)},C_{\mathrm{out}}^{(2)},C_{\mathrm{out}}^{(3)})=(c_1,c_2,c_3)$.

\subsection{Student architectures}
\label{sec:supp_student_arch}

The PQKD student matches the teacher macro-architecture (same block structure and GAP head) but replaces selected convolutional layers by photonic-conditioned dictionary convolutions (Supplementary Note~\ref{sec:supp_pcdconv}). Concretely, for each compressed convolution layer $\ell\in\mathcal{L}$, the dense kernel is replaced by a trainable spatial basis $S^{(\ell)}\in\R^{R_\ell\times k_\ell\times k_\ell}$ and a photonic-generated mixing tensor $M^{(\ell)}$:
\begin{equation}
\mathrm{vec}(M^{(\ell)}) = A^{(\ell)} z(\theta),\qquad
W^{(\ell)}_{o,i,:,:}=\sum_{r=1}^{R_\ell} M^{(\ell)}_{o,i,r}\,S^{(\ell)}_{r,:,:},
\qquad \ell\in\mathcal{L}.
\end{equation}
Layers not in $\mathcal{L}$ remain standard dense convolutions with fully trainable kernels. We evaluate three compression configurations (Supplementary Note~\ref{sec:supp_compression_configs}): compress \texttt{conv1} only; compress \texttt{conv1+conv2}; and compress all convolution layers.

\subsection{Baselines and controls}
\label{sec:supp_baselines}

Because PQKD combines two ingredients---structured parameterisation and a photonic conditioning signal---we include baseline models that isolate each factor. The goal is to distinguish: (i) gains attributable to distillation alone, (ii) gains attributable to dictionary structure alone, and (iii) gains attributable specifically to photonic-conditioned parameter generation.

\begin{enumerate}
\item \textbf{KD-only student (no compression).}
A student with reduced width (or identical architecture) trained using the distillation objective in Eq.~\eqref{eq:kd_loss_supp}, without replacing any convolution kernels. This baseline quantifies the benefit of knowledge distillation absent structured compression.

\item \textbf{Dictionary-only student (no photonic conditioning).}
Replace the target convolution layers by the same dictionary decomposition
$W_{o,i}=\sum_{r} M_{o,i,r}S_{r}$, but treat $M$ as a \emph{trainable} parameter tensor rather than generating it from $z(\theta)$. This isolates the effect of the dictionary factorisation while removing the photonic control constraint.

\item \textbf{Random-feature conditioning.}
Replace the photonic feature $z(\theta)$ by i.i.d.\ Gaussian features $\tilde z\sim\mathcal{N}(0,I_d)$ (optionally re-sampled per epoch or fixed per run), while keeping the same fixed projection $A$ and dictionary basis parameterisation. This baseline tests whether PQKD improvements could be explained by generic stochastic conditioning rather than photonic structure.

\item \textbf{Fixed photonic feature (no $\theta$ learning).}
Use the photonic sampling and feature map to compute $z(\theta)$ but keep $\theta$ fixed throughout training (no SPSA or other updates). This baseline isolates the value of \emph{learning} $\theta$ as opposed to using a static photonic feature.
\end{enumerate}

When reporting results, we keep training budgets (epochs, batch sizes, optimiser settings) identical across PQKD and baselines unless explicitly stated, and we use the same teacher checkpoint for all student variants in a given experimental run.

\subsection{Parameter counting conventions}
\label{sec:supp_param_conventions}

We report compression ratios under two complementary conventions to avoid ambiguity:

\begin{itemize}
\item \textbf{Overall trainable-parameter ratio:}
\begin{equation}
\mathrm{CR}_{\mathrm{overall}} = 
\frac{\#\mathrm{params}_{\mathrm{teacher}}}{\#\mathrm{params}_{\mathrm{student}}}
\quad\text{with $\#\mathrm{params}_{\mathrm{student}}$ including }\dim(\theta)\text{ once.}
\end{equation}

\item \textbf{Conv-only ratio:}
\begin{equation}
\mathrm{CR}_{\mathrm{conv}} = 
\frac{\#\mathrm{params}_{\mathrm{teacher,conv}}}{\#\mathrm{params}_{\mathrm{student,conv}}},
\end{equation}
computed by restricting the parameter count to convolutional layers only. This isolates compression achieved in the feature extractor independent of the classifier head.
\end{itemize}

Fixed projection matrices $A^{(\ell)}$ are excluded from all trainable-parameter counts because they are not optimised; however, they may contribute to memory footprint and can be implemented in structured form if required. Photonic parameters $\theta$ are included in the student trainable budget by default (unless explicitly stated otherwise) to reflect the full learnable budget of the PQKD mechanism.

% =========================================================
\section{Supplementary Note 7: Quantum model details}
\label{sec:sn7}

\subsection{Circuit family and theta-length constraint}
\label{sec:supp_theta_constraint}

Our PQKD implementation interfaces with a programmable CV photonic sampler whose native parameterisation is tied to a specific interferometric circuit family. In particular, the backend adopts a \emph{tiled} interferometer structure consistent with loop- or time-bin architectures, where each ``tiling'' (or layer) implements a fixed-pattern mode mixing described by a sequence of beam-splitter operations interleaved with phase shifts. Within this circuit family, the number of independent mixing angles per tiling scales linearly with the number of modes $N$, rather than quadratically as in a fully general $U(N)$ mesh. This architectural choice reflects realistic constraints of compact photonic implementations (e.g., single-loop interferometers), and it directly determines the admissible length of the continuous parameter vector $\theta$ accepted by the sampler.

\paragraph{Per-tiling parameter count.}
For an $N$-mode instance, a single tiling is parameterised by
\begin{equation}
L_{\mathrm{tile}} = N-1
\end{equation}
real-valued angles associated with beam-splitter reflectivities (and/or equivalent mixing parameters) in the fixed connectivity pattern of the backend. For the main experiments with $N=16$ modes, this implies
\begin{equation}
L_{\mathrm{tile}} = N-1 = 15.
\end{equation}
We denote by $\theta^{(t)}\in\R^{L_{\mathrm{tile}}}$ the parameter vector controlling tiling $t$.

\paragraph{Tiling interpretation for extended parameter vectors.}
To support systematic sweeps over the effective photonic parameter budget $\dim(\theta)$, we allow user-specified vectors whose length may exceed $L_{\mathrm{tile}}$. When $\dim(\theta)>L_{\mathrm{tile}}$, the backend interprets the circuit as a concatenation of multiple identical connectivity tiles, each with its own set of $(N-1)$ parameters. Specifically, we define the number of tiles as
\begin{equation}
n_{\mathrm{tiling}}=\left\lceil \frac{\dim(\theta)}{N-1}\right\rceil,
\label{eq:ntiling_def}
\end{equation}
so that the circuit consumes a total of
\begin{equation}
L_{\mathrm{eff}}=(N-1)\,n_{\mathrm{tiling}}
\label{eq:Leff_def}
\end{equation}
angles. Conceptually, this corresponds to composing the interferometer as a product
\begin{equation}
U(\theta_{\mathrm{fit}})=U^{(n_{\mathrm{tiling}})}(\theta^{(n_{\mathrm{tiling}})})\cdots U^{(2)}(\theta^{(2)})U^{(1)}(\theta^{(1)}),
\end{equation}
where each $U^{(t)}$ is a single-tiling interferometric transformation parameterised by $\theta^{(t)}\in\R^{N-1}$.

\paragraph{Deterministic adjustment for sampler compatibility.}
Because the sampler expects exactly $L_{\mathrm{eff}}$ parameters, we map the user-specified $\theta\in\R^{\dim(\theta)}$ to a backend-compatible vector
\begin{equation}
\theta \mapsto \theta_{\mathrm{fit}} \in \R^{(N-1)\,n_{\mathrm{tiling}}},
\label{eq:theta_fit_space}
\end{equation}
using a deterministic rule that preserves reproducibility:
\begin{equation}
\theta_{\mathrm{fit}}=
\begin{cases}
\big[\theta;\, \mathbf{0}\big] & \text{if } \dim(\theta) < L_{\mathrm{eff}} \quad \text{(zero-padding)},\\[4pt]
\theta_{1:L_{\mathrm{eff}}} & \text{if } \dim(\theta) > L_{\mathrm{eff}} \quad \text{(truncation)},\\[4pt]
\theta & \text{if } \dim(\theta) = L_{\mathrm{eff}}.
\end{cases}
\label{eq:theta_fit_rule}
\end{equation}
Here $[\theta;\mathbf{0}]$ denotes concatenation with zeros. This adjustment guarantees that every call to the photonic sampler receives a valid parameter vector, while allowing us to treat $\dim(\theta)$ as a controlled hyperparameter in compression sweeps.

\paragraph{Implementation details and practical considerations.}
For numerical stability and physical interpretability of the underlying interferometric angles, we additionally apply a fixed bounding operation to $\theta_{\mathrm{fit}}$ when required by the backend (e.g., wrapping angles to a principal interval):
\begin{equation}
\theta_{\mathrm{fit}} \leftarrow \mathrm{wrap}(\theta_{\mathrm{fit}};[-\pi,\pi]) \quad \text{or} \quad \theta_{\mathrm{fit}} \leftarrow \mathrm{clip}(\theta_{\mathrm{fit}};[-\theta_{\max},\theta_{\max}]),
\end{equation}
where the specific convention is held constant across all runs. Importantly, the tiling expansion in Eq.~\eqref{eq:ntiling_def} increases the effective circuit depth (number of repeated mixing stages) without changing the mode count $N$, and therefore provides a principled way to scale the photonic control capacity while maintaining a fixed feature dimension $d$ and a fixed sampling interface.

Overall, this theta-length constraint is not merely a software artefact: it reflects a physically motivated parameterisation of a structured interferometric circuit family. Our deterministic mapping in Eq.~\eqref{eq:theta_fit_rule} ensures strict sampler compatibility, reproducible hyperparameter sweeps over $\dim(\theta)$, and a clear interpretation of additional photonic parameters as additional interferometric tiles.
\subsection{Measurement and sample format}
\label{sec:supp_measurement_format}

PQKD accesses the photonic processor through repeated measurements. Each circuit execution (shot) produces a multi-mode detection outcome whose statistics depend on the circuit parameters $\theta$ and the chosen input state preparation. To make the downstream feature construction well-defined and hardware-aligned, we adopt a standard \emph{thresholded} measurement representation that converts raw detection events into a fixed-length binary vector over modes.

\paragraph{Measurement model.}
Let $\rho_\theta = U(\theta)\rho_{\mathrm{in}}U(\theta)^\dagger$ be the output state of the $N$-mode photonic circuit. A single shot applies a measurement described by a POVM $\{E_\omega\}_{\omega\in\Omega}$ and returns a random outcome $\omega\sim p_\theta(\omega)=\Tr(E_\omega\rho_\theta)$. In photon-counting settings\cite{hadfield2009single}, $\omega$ can be represented as a photon-number pattern
\begin{equation}
n = (n_1,\ldots,n_N)\in\mathbb{Z}_{\ge 0}^{N},
\end{equation}
where $n_j$ is the number of detected photons in mode $j$. In threshold-detection settings, the detector reports only whether each mode is ``clicked'' (occupied) or ``not clicked'' (empty)\cite{quesada2018gaussian}, which can be viewed as a coarse-graining of photon-number measurements.

\paragraph{Thresholding to a binary occupancy vector.}
For consistency across backends and to reduce sensitivity to rare high-count events, we convert each raw pattern $n$ into a binary occupancy vector
\begin{equation}
b = (b_1,\ldots,b_N)\in\{0,1\}^{N},
\qquad
b_j = \mathbb{I}[\,n_j>0\,],
\label{eq:thresholding}
\end{equation}
where $\mathbb{I}[\cdot]$ is the indicator function. Thus, each shot yields a length-$N$ bitstring $b$ encoding per-mode occupancy. We denote the resulting random variable by $B\sim P_\theta$ over $\{0,1\}^N$, where $P_\theta$ is the induced distribution after thresholding.

\paragraph{Finite-shot sampling and empirical distribution.}
For a single feature evaluation, we collect $S$ independent samples
\begin{equation}
b^{(1)},\ldots,b^{(S)} \ \overset{\mathrm{i.i.d.}}{\sim}\ P_\theta,
\label{eq:S_samples}
\end{equation}
and compute empirical statistics (e.g., histograms of selected marginals) to form $\widehat{p}_\theta$ and the corresponding feature vector $z(\theta)=\Phi(\widehat{p}_\theta)$ (Supplementary Note~\ref{sec:supp_feature_map}). In all experiments, $S$ is treated as an explicit hyperparameter controlling the trade-off between feature variance and evaluation cost, and the impact of finite-shot noise is studied in Supplementary Note~\ref{sec:supp_finite_shot} and Supplementary Note~\ref{sec:exp3}.

\paragraph{Implementation.}
In code, the sampler returns either an array of per-shot samples of shape $(S,N)$ with entries in $\{0,1\}$ (after thresholding), or an equivalent representation (e.g., integer-encoded bitstrings) that can be deterministically converted to the binary format in Eq.~\eqref{eq:thresholding}. We standardise all outputs to the binary matrix representation to ensure that feature construction and parameter generation are backend-agnostic.

\subsection{Feature construction}
\label{sec:supp_feature_512}

This subsection specifies the exact photonic feature map $\Phi$ used in the main experiments. The guiding design goal is to transform shot-limited binary detection patterns into a \emph{fixed-length}, \emph{order-consistent}, and \emph{variance-controlled} representation that (i) is simple to implement, (ii) remains stable under finite-shot sampling, and (iii) preserves meaningful distributional information beyond per-mode means.

\paragraph{Bitstring partitioning.}
Each shot produces a thresholded occupancy vector $b\in\{0,1\}^{N}$ with $N=16$ modes (Supplementary Note~\ref{sec:supp_measurement_format}). We partition the bitstring into two contiguous 8-bit halves,
\begin{equation}
b=\big[b^{(1)};\,b^{(2)}\big],\qquad
b^{(1)}\in\{0,1\}^{8},\ \ b^{(2)}\in\{0,1\}^{8}.
\label{eq:bit_split}
\end{equation}
This produces two low-dimensional views (marginals) of the $16$-bit outcome. Importantly, this construction avoids the exponential $(2^{16})$ histogram that would be too sparse under realistic shot budgets, while still capturing structured multi-bit correlations within each half.

\paragraph{Index mapping.}
For each half $b^{(m)}=(b^{(m)}_1,\ldots,b^{(m)}_8)$, we define a deterministic integer encoding
\begin{equation}
\mathrm{idx}\!\left(b^{(m)}\right)
=
\sum_{j=1}^{8} b^{(m)}_j\,2^{8-j}
\in\{0,1,\ldots,255\},
\qquad m\in\{1,2\},
\label{eq:idx_def}
\end{equation}
where the bit ordering convention (most-significant to least-significant) is fixed throughout all runs. Any consistent convention is valid as long as it is held constant across training and evaluation.

\paragraph{Empirical marginal histograms.}
Given $S$ i.i.d.\ samples $\{b_s\}_{s=1}^{S}$, we compute the two empirical marginal histograms $h^{(1)},h^{(2)}\in\R^{256}$:
\begin{equation}
h^{(m)}[k]
=
\frac{1}{S}\sum_{s=1}^{S}\mathbf{1}\Big\{\mathrm{idx}\!\left(b^{(m)}_s\right)=k\Big\},
\qquad k\in\{0,\ldots,255\},\ \ m\in\{1,2\},
\label{eq:hist_def}
\end{equation}
where $\mathbf{1}\{\cdot\}$ is the indicator function. By construction, each histogram is a valid empirical probability mass function:
\begin{equation}
h^{(m)}[k]\ge 0,\qquad \sum_{k=0}^{255} h^{(m)}[k]=1.
\label{eq:hist_norm}
\end{equation}

\paragraph{Concatenation and standardisation.}
We form a $512$-dimensional feature by concatenating the two marginals:
\begin{equation}
\tilde z(\theta)=\big[h^{(1)};\,h^{(2)}\big]\in\R^{512}.
\label{eq:z_concat}
\end{equation}
To improve numerical conditioning and to make the scale of the conditioning signal comparable across runs, we apply an affine standardisation:
\begin{equation}
z(\theta)
=
\gamma\cdot
\frac{\tilde z(\theta)-\mu}{\sigma+\epsilon}
\in\R^{512},
\label{eq:z_standardise}
\end{equation}
where $\mu\in\R^{512}$ and $\sigma\in\R^{512}$ are fixed reference statistics and $\epsilon>0$ is a small stabiliser (added elementwise) to avoid division by zero. In our implementation, $(\mu,\sigma)$ are computed once from an initial set of feature evaluations (e.g., using a fixed seed and the initial $\theta$) and then held fixed for the remainder of training; this ensures that the mapping $\Phi$ remains deterministic and does not leak label information.

\paragraph{Interpretation and trade-offs.}
The two-marginal design in Eq.~\eqref{eq:hist_def}--\eqref{eq:z_standardise} offers a practical compromise between expressivity and sample efficiency. A full histogram over $\{0,1\}^{16}$ would require $2^{16}=65{,}536$ bins and would be extremely sparse for typical shot budgets. In contrast, the two 8-bit marginals use only $2\times 256=512$ bins, yielding more reliable estimates under finite shots while preserving higher-order correlations within each half. The finite-shot concentration behaviour follows the scaling in Supplementary Note~\ref{sec:supp_finite_shot} with effective histogram size $K=512$.

\paragraph{Implementation note.}
In code, if samples are returned as a binary matrix $B\in\{0,1\}^{S\times 16}$, the index mapping in Eq.~\eqref{eq:idx_def} can be computed by a dot product with a fixed weight vector $(2^{7},2^{6},\ldots,2^{0})$. Histograms are then obtained by counting occurrences over $k\in\{0,\ldots,255\}$ and normalising by $S$. The final $z(\theta)$ is produced by concatenation and elementwise standardisation as in Eq.~\eqref{eq:z_standardise}.

% =========================================================
\section{Supplementary Note 8: Training protocol}
\label{sec:sn8}

This Supplementary Note describes the full training procedure used for PQKD, including teacher optimisation, student optimisation under knowledge distillation, and the update rule for photonic parameters. The protocol is designed to be reproducible and to reflect the operational constraints of a shot-based photonic backend: the photonic module is accessed through repeated sampling to construct features, while the classical student is trained using standard gradient-based optimisation. Because the photonic feature affects the student through a non-analytic sampling interface, we adopt an alternating optimisation strategy that approximates a bilevel objective.

\subsection{Data splits, preprocessing, and evaluation}
\label{sec:supp_data_eval_protocol}

Unless otherwise stated, we use a fixed train/validation/test split and fixed random seeds for dataset subsampling and mini-batch ordering. Inputs are standardised according to dataset conventions (e.g., per-channel normalisation for CIFAR-10) and augmented only when explicitly indicated. All reported validation and test metrics are computed deterministically (no dropout, fixed batch ordering, no data augmentation at evaluation time). Accuracy is reported as top-1 classification accuracy; cross-entropy refers to the standard negative log-likelihood loss.

\subsection{Teacher training}
\label{sec:supp_teacher_training}

Teachers are trained with hard-label cross-entropy using Adam with default learning rate $10^{-3}$ for 100 epochs unless otherwise stated. We use batch size 64 and apply standard regularisation consistent with the teacher architecture (e.g., dropout in \texttt{conv3}). Let teacher logits be $t(x)$ and hard labels be $y$. The teacher objective is
\begin{equation}
\min_{\psi}\ \E_{(x,y)\sim\mathcal{D}_{\mathrm{train}}}\Big[\CE\big(y,\softmax(t(x;\psi))\big)\Big],
\end{equation}
where $\psi$ denotes teacher parameters. The best-performing checkpoint is selected using validation accuracy and then evaluated once on the held-out test set.

\subsection{Alternating optimisation for PQKD}
\label{sec:supp_alt_opt}

\paragraph{Bilevel viewpoint.}
PQKD introduces two coupled sets of learnable variables: classical student weights $w$ and photonic circuit parameters $\theta$. The student weights are optimised to minimise a training distillation loss, while $\theta$ is optimised to improve generalisation as measured by a validation proxy objective computed using the photonic feature map. This structure can be expressed as the bilevel problem
\begin{equation}
w^{\ast}(\theta)\in\arg\min_{w}\ \mathcal{L}^{\mathrm{train}}_{\mathrm{KD}}(w,\theta),
\qquad
\theta^{\ast}\in\arg\min_{\theta}\ \mathcal{L}^{\mathrm{val}}_{\mathrm{KD}}\!\big(w^{\ast}(\theta),\theta\big).
\label{eq:bilevel}
\end{equation}
Direct bilevel optimisation is typically computationally expensive, especially when the outer variable $\theta$ only influences the model through shot-based samples. We therefore employ an alternating scheme that approximates Eq.~\eqref{eq:bilevel} with a small number of outer updates per epoch.

\paragraph{Operational training loop.}
Each student epoch consists of two phases:
\begin{enumerate}
\item \textbf{Photonic phase (outer updates):} perform $\texttt{theta\_updates\_per\_epoch}=10$ SPSA steps to update $\theta$ while holding $w$ fixed. Each SPSA step uses a validation mini-batch (or a small fixed set of validation batches) to evaluate the proxy objective $J(\theta)$.
\item \textbf{Classical phase (inner updates):} perform one epoch of Adam updates on $w$ using the distillation loss, holding $\theta$ fixed (equivalently, holding the photonic feature used by the student fixed for that epoch).
\end{enumerate}
This schedule yields a practical and stable separation of concerns: classical optimisation proceeds with smooth gradients on training data, while photonic optimisation is driven by a generalisation-oriented signal from validation performance and uses only function evaluations through the sampling backend.

\subsection{SPSA photonic updates}
\label{sec:supp_spsa}

\paragraph{Validation proxy objective.}
Let $J(\theta)$ denote the scalar objective used for outer-loop updates. In our implementation, $J(\theta)$ is defined as the negative validation accuracy (or, equivalently, the validation distillation loss) computed with the current student weights $w$ and with the photonic feature map evaluated at $\theta$. Because the feature is shot-limited, $J(\theta)$ is stochastic; we reduce variance by averaging over a small number of validation mini-batches.

\paragraph{SPSA estimator.}
Simultaneous Perturbation Stochastic Approximation (SPSA) provides an efficient gradient-free update that requires only two objective evaluations per step, independent of $\dim(\theta)$\cite{spall2002multivariate}. At SPSA step $k$, we draw a Rademacher perturbation vector
\begin{equation}
\Delta_k \in \{\pm 1\}^{\dim(\theta)}
\qquad\text{with i.i.d.\ entries}.
\end{equation}
We then evaluate the objective at symmetric perturbations
\begin{equation}
\theta_k^{\pm} = \theta_k \pm c\,\Delta_k,
\end{equation}
and form the stochastic gradient estimate
\begin{equation}
\widehat{g}_k
=
\frac{J(\theta_k^{+})-J(\theta_k^{-})}{2c}\,\Delta_k.
\label{eq:spsa_grad}
\end{equation}
The parameter update is
\begin{equation}
\theta_{k+1}=\theta_k-a\,\widehat{g}_k,
\label{eq:spsa_update}
\end{equation}
where $a>0$ is the step size and $c>0$ is the perturbation scale. For stability and to respect backend parameter constraints, we apply elementwise clipping
\begin{equation}
\theta_{k+1} \leftarrow \mathrm{clip}(\theta_{k+1};[-\theta_{\max},\theta_{\max}]).
\end{equation}
In all experiments, $(a,c,\theta_{\max})$ are held fixed within a sweep and reported alongside the corresponding results.

\paragraph{Relation to finite-shot noise.}
Because $J(\theta)$ depends on shot-limited features (Supplementary Note~\ref{sec:supp_finite_shot}), $\widehat{g}_k$ is a noisy estimate. The use of symmetric perturbations in Eq.~\eqref{eq:spsa_grad} cancels some common-mode sampling noise, and evaluating $J(\theta)$ on multiple validation mini-batches further reduces variance. We report noise impact studies by varying the shot budget $S$ and observing its effect on stability and final accuracy (Supplementary Note~\ref{sec:supp_ema}).

\subsection{Classical student updates}
\label{sec:supp_student_updates}

During the classical phase, student weights $w$ are optimised using Adam on the training distillation objective. Let $s(x;w,\theta)$ denote student logits given the current photonic feature (fixed for the epoch). The update minimises $\mathcal{L}^{\mathrm{train}}_{\mathrm{KD}}(w,\theta)$ defined in Eq.~\eqref{eq:kd_loss_supp}. We use the same batch size (64) as the teacher unless otherwise stated. Training is performed with standard best practices: gradients are backpropagated only through the classical student network; no gradients are propagated through the photonic sampler, consistent with our black-box sampling interface.
\subsection{EMA feature smoothing}
\label{sec:supp_ema}

Finite-shot photonic sampling produces a stochastic estimate of the conditioning feature $z(\theta)$ used to parameterise the compressed student. If injected directly, this sampling noise appears as high-frequency, epoch-to-epoch innovations in the feature stream and can be amplified by the student’s sensitivity to $z$, degrading optimisation stability at fixed shot budgets (Supplementary Note~\ref{sec:supp_finite_shot}). To suppress these fluctuations without modifying the photonic circuit or measurement procedure, we optionally apply exponential moving average (EMA) smoothing to the feature sequence:
\begin{equation}
\bar z_t \;=\; \beta\,\bar z_{t-1} + (1-\beta)\,z_t(\theta),
\qquad \beta\in[0,1),
\label{eq:ema_feature}
\end{equation}
where $t$ indexes feature evaluations (epochs in our implementation) and $\bar z_0 = z_0(\theta)$. When enabled, the student forward pass uses $\bar z_t$ in place of the instantaneous sample $z_t(\theta)$.

EMA acts as a causal low-pass filter: it preserves slowly varying components of the photonic signal while attenuating rapid innovations that are predominantly induced by finite-shot fluctuations. A useful interpretation follows from the decomposition $z_t(\theta)=\mu(\theta)+\varepsilon_t$ with $\mathbb{E}[\varepsilon_t]=0$ and approximately independent increments $\varepsilon_t$ across successive evaluations. Under this standard approximation, $\mathbb{E}[\bar z_t]=\mu(\theta)$ (no bias in the mean signal), and the variance of the smoothed estimate is reduced relative to the raw estimate by
\begin{equation}
\frac{\mathrm{Var}(\bar z_t)}{\mathrm{Var}(z_t)} \;\approx\; \frac{1-\beta}{1+\beta},
\label{eq:ema_var_ratio}
\end{equation}
with the reduction becoming sharper as $\beta\rightarrow 1$. Equivalently, EMA yields an effective increase in sampling budget,
\begin{equation}
S_{\mathrm{eff}} \;\approx\; \frac{1+\beta}{1-\beta}\,S,
\label{eq:ema_seff}
\end{equation}
so that the student experiences a feature stream comparable to one obtained with substantially more shots, but at the same physical measurement cost per epoch.

In our MNIST EMA-mechanism trace at $S{=}200$ shots and $\beta{=}0.9$, the distribution of per-dimension variance ratios $\mathrm{Var}(z_{\mathrm{used}})/\mathrm{Var}(z_{\mathrm{raw}})$ concentrates well below unity, with a median $0.0406$ (IQR $0.0322$--$0.0532$) over the analysis window, consistent with the theoretical attenuation level $(1-\beta)/(1+\beta)=0.0526$ and the corresponding amplification $S_{\mathrm{eff}}\approx 19\,S$. These observations support the role of EMA as a simple, hardware-compatible variance-reduction mechanism for stabilising PQKD in shot-limited regimes, and complement the robustness results in Sec.~\ref{sec:exp3}.

\begin{figure}[t]
  \centering
  \includegraphics[width=0.95\linewidth]{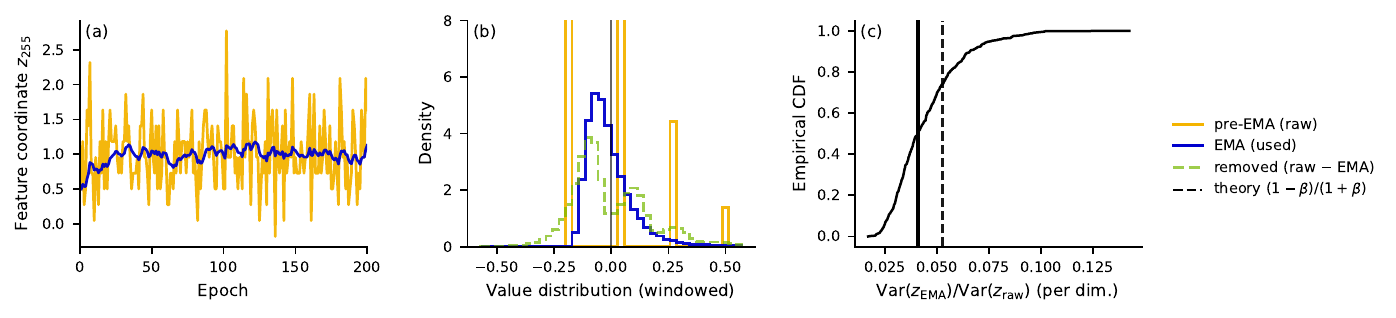}
  \caption{\textbf{EMA suppresses shot-noise fluctuations in the photonic conditioning signal.}
  Feature traces recorded during PQKD training at fixed shot budget compare the raw per-epoch photonic feature (pre-EMA) with the EMA-smoothed feature used by the student.
  \textbf{(a)} A representative feature coordinate illustrates attenuation of high-frequency sampling jitter under EMA.
  \textbf{(b)} Windowed value distributions across feature dimensions for the raw and EMA-used signals; the dashed curve indicates the removed residual $(z_{\mathrm{raw}}-z_{\mathrm{EMA}})$.
  \textbf{(c)} Empirical CDF of the per-dimension variance ratio $\mathrm{Var}(z_{\mathrm{EMA}})/\mathrm{Var}(z_{\mathrm{raw}})$ with the expected attenuation level shown for reference.}
  \label{fig:ema_mechanism}
\end{figure}

\subsection{Algorithms}

Algorithms~\ref{alg:pqkd_train}--\ref{alg:feature} summarise the full PQKD training loop, the SPSA outer-loop update, and the photonic feature extraction used in the main experiments ($N=16$, $d=512$). These procedures directly implement the alternating optimisation described in Eq.~\eqref{eq:bilevel} and Eq.~\eqref{eq:spsa_update}.

\begin{algorithm}[b]
\caption{PQKD training loop (teacher $\rightarrow$ student + photonic updates)}
\label{alg:pqkd_train}
\begin{algorithmic}[1]
\Require Data loaders $\mathcal{D}_{\mathrm{train}},\mathcal{D}_{\mathrm{val}}$, teacher $f_T$, student $f_S$, photonic sampler $\mathsf{Sampler}$,
KD hyperparameters $(\tau,\lambda)$, shots $S$, outer steps $U_{\mathrm{out}}$, student epochs $E_{\mathrm{student}}$
\State Train teacher $f_T$ with cross-entropy on $\mathcal{D}_{\mathrm{train}}$; freeze $f_T$
\State Initialise photonic parameters $\theta\leftarrow 0$; initialise student parameters $w$
\State Initialise (optional) EMA state $\bar z \leftarrow \emptyset$
\For{$\mathrm{epoch}=1$ to $E_{\mathrm{student}}$}
  \Comment{\textbf{Photonic phase: update $\theta$ with $w$ fixed}}
  \For{$u=1$ to $U_{\mathrm{out}}$}
    \State $\theta \leftarrow \textsc{SPSAUpdate}(\theta, w, f_T, f_S, \mathcal{D}_{\mathrm{val}}, \mathsf{Sampler}, S, \tau,\lambda)$
    \Comment{Algorithm~\ref{alg:spsa}}
  \EndFor
  \Comment{\textbf{Classical phase: update $w$ with $\theta$ fixed}}
  \State $z(\theta) \leftarrow \textsc{PhotonicFeature}(\mathsf{Sampler},\theta,S)$
  \Comment{Algorithm~\ref{alg:feature}}
  \If{EMA enabled}
     \State $\bar z \leftarrow \beta \bar z + (1-\beta) z(\theta)$ \ \ (initialise $\bar z=z(\theta)$ if empty)
     \State Use $\bar z$ as the conditioning feature for this epoch
  \Else
     \State Use $z(\theta)$ as the conditioning feature for this epoch
  \EndIf
  \State Train student for one epoch with Adam on $\mathcal{L}_{\mathrm{KD}}(w,\theta)$ using $\mathcal{D}_{\mathrm{train}}$
\EndFor
\end{algorithmic}
\end{algorithm}

\begin{algorithm}[b]
\caption{SPSA update for photonic parameters}
\label{alg:spsa}
\begin{algorithmic}[1]
\Require Current $\theta$, fixed student weights $w$, step sizes $(a,c)$, clipping bound $\theta_{\max}$,
validation objective $J(\theta)$ computed from KD loss or $-\,$accuracy on $\mathcal{D}_{\mathrm{val}}$
\State Sample $\Delta\in\{\pm1\}^{\dim(\theta)}$
\State $\theta^{+} \leftarrow \mathrm{clip}(\theta + c\Delta,\ -\theta_{\max},\theta_{\max})$
\State $\theta^{-} \leftarrow \mathrm{clip}(\theta - c\Delta,\ -\theta_{\max},\theta_{\max})$
\State $J^{+} \leftarrow J(\theta^{+})$ \Comment{evaluate using photonic feature extraction + KD forward on $\mathcal{D}_{\mathrm{val}}$}
\State $J^{-} \leftarrow J(\theta^{-})$
\State $\widehat{g}\leftarrow \dfrac{J^{+}-J^{-}}{2c}\Delta$
\State $\theta \leftarrow \mathrm{clip}(\theta - a\widehat{g},\ -\theta_{\max},\theta_{\max})$
\State \Return $\theta$
\end{algorithmic}
\end{algorithm}

\begin{algorithm}[!t]
\caption{Photonic feature extraction ($N=16$, $d=512$)}
\label{alg:feature}
\begin{algorithmic}[1]
\Require Photonic sampler $\mathsf{Sampler}$, parameters $\theta$, shots $S$, fixed standardisation $(\mu,\sigma,\epsilon)$, scaling $\gamma$
\State Sample $S$ outcomes over $N=16$ modes; threshold to bits $b_s\in\{0,1\}^{16}$
\State Split each $b_s$ into two halves $b_s^{(1)},b_s^{(2)}\in\{0,1\}^8$; map each half to $\mathrm{idx}\in\{0,\dots,255\}$
\State Form two normalised histograms $h^{(1)},h^{(2)}\in\mathbb{R}^{256}$ and concatenate $\tilde z \leftarrow [h^{(1)};h^{(2)}]\in\mathbb{R}^{512}$
\State Standardise and scale: $z(\theta)\leftarrow \gamma \cdot \dfrac{\tilde z-\mu}{\sigma+\epsilon}$
\State \Return $z(\theta)$
\end{algorithmic}
\end{algorithm}

\end{document}